\documentclass[twocolumn,showpacs,showkeys,preprintnumbers,floatfix,letterpaper,prc]{revtex4}

\usepackage{citesort,enumerate}
\usepackage{amsmath,amssymb}  
\usepackage{bm}               
\usepackage{amscd}            
\usepackage{graphicx}         
\usepackage{hhline,multirow}  
\usepackage{dcolumn}          

\newcommand{\beq}{\begin{equation}}
\newcommand{\eeq}{\end{equation}}
\newcommand{\bea}{\begin{eqnarray}}
\newcommand{\beas}{\begin{eqnarray*}}
\newcommand{\beau}[1]{\begin{equation} \label{#1} \begin{array}{rcl}}
\newcommand{\eea}{\end{eqnarray}}
\newcommand{\eeas}{\end{eqnarray*}}
\newcommand{\eeau}{\end{array} \end{equation}}
\newcommand{\bay}{\begin{array}}
\newcommand{\eay}{\end{array}}
\newcommand{\bals}{\begin{align*}}
\newcommand{\eals}{\end{align*}}

\newcommand{\lora}{{\longrightarrow}}

\newcommand{\ra}{{\rightarrow}}

\newcommand{\bra}[1]{\langle #1 \vert}
\newcommand{\ket}[1]{\vert #1 \rangle}

\newcommand{\vev}[1]{\langle #1 \rangle}

\newcommand{\Tr}{\mbox{Tr}}

\newcommand{\nbar}{{\overline n}}

\newcommand{\ksl}{k\!\!\!\slash} 
\newcommand{\lsl}{l\!\!\slash} 
\newcommand{\nsl}{n\!\!\!\slash} 
\newcommand{\nbsl}{\overline n\!\!\!\slash}

\newcommand{\HH}{{\mathcal H}}

\newcommand{\sth}{s_\text{th}}
\begin{document}


\preprint{JLAB-THY-08-828}

\title{Collinear factorization for deep inelastic scattering 
structure functions \\ at large Bjorken $\bm x_B$}

\author{Alberto~Accardi$^{a,b,c}$ and Jian-Wei~Qiu$^c$}
\affiliation{
$^a$Hampton University, Hampton, VA, 23668, USA \\
$^b$Jefferson Lab, Newport News, VA 23606, USA \\
$^c$Department of Physics and Astronomy, Iowa State University,
Ames, IA 50011-3160, USA 
}

\begin{abstract}
We examine the uncertainty of perturbative QCD factorization for 
hadron structure functions in deep inelastic scattering at a large 
value of the Bjorken variable $x_B$.  
We analyze the target mass correction to the
structure functions by using the 
collinear factorization
approach in the momentum space.  
We express the long distance physics of structure 
functions and the leading target mass corrections in terms of
parton distribution functions with the standard operator 
definition.  We compare our result with existing work on the 
target mass correction.  We also discuss the impact of 
a final-state jet function on the extraction of parton distributions 
at large fractional momentum $x$.
\end{abstract}

\pacs{}
\keywords{}

\maketitle


\section{Introduction}
\label{Intro}

Much of the predictive power of perturbative Quantum Chromo Dynamics 
(pQCD) is contained in factorization theorems and in the universality 
of non-perturbative hadronic matrix elements \cite{Collins:1989gx}.  
Predictions follow when processes with different hard scatterings 
but the same matrix elements are compared.  
In the case of leading power contributions, the universal matrix
elements are interpreted as parton (quark or gluon) distribution
functions (PDFs).  With the PDFs extracted from a global QCD 
analysis \cite{Pumplin:2002vw,Martin:2002aw,Gluck:1998xa},
pQCD has been very successful in interpreting and predicting 
high-energy scattering processes. 

However, significant uncertainties still exist in the PDFs
due to the accuracy of experimental data and to unknown higher order 
corrections to perturbative calculations.  In particular, the PDFs
are least constrained in the region where the parton momentum 
fraction $x>0.5$ for valence quark distributions and $x>0.3$ for
gluon and sea quark distributions \cite{Pumplin:2002vw,Martin:2002aw}.  
On the other hand, 
precise PDFs are needed for many reasons \cite{Kuhlmann:1999sf}.  
For example, the discovery potential of the 
Large Hadron Collider (LHC) on new physics as excess 
in particle/jet spectrum at large momentum requires 
accurate PDFs at large $x$ and large factorization scale $\mu$.  
Since PDFs at a large $\mu$ are obtained by solving DGLAP evolution 
equations with input PDFs at a lower factorization scale, and 
the evolution feeds the large-$x$ partons at a lower scale to 
those at a higher scale with smaller momentum fraction $x$, 
the precision of PDFs at large $\mu$ depends on the accuracy 
of PDFs at large $x$ and low factorization scale.
Furthermore, reliable information on the ratio of 
$d(x)/u(x)$ as $x\rightarrow 1$ could provide very important insights
into the non perturbative structure of the nucleon
\cite{Melnitchouk:1995fc,Brodsky:1994kg,Isgur:1998yb,Farrar:1975yb} 
and references therein. 
However, because of the PDFs steeply falling shape as a function of
$x$ as $x\ra 1$, and because of the convolution of two PDFs,
most observables in hadronic collisions do not provide tight enough 
constraints to the PDFs at large $x$.  On the other hand, inclusive 
lepton-hadron deep inelastic scattering (DIS) at large Bjorken $x_B$ 
is a more direct and clean probe of large-$x$ parton distributions.
Recently, experiments at the Jefferson Laboratory have produced 
DIS data at large $x_B$ with high precision, but at relatively 
low virtuality $Q^2$ of the exchanged virtual photon in 
lepton-hadron collisions 
\cite{Tvaskis:2006tv,Melnitchouk:2004ep,Cardman:2006xt}.
Experiments measure DIS  
cross sections, or, equivalently, the DIS structure functions, not PDFs.
In order to extract PDFs at large $x$ from these and other data at low
$Q^2$, it is necessary to have theoretical control over power
corrections,  such as the dynamical power corrections (or high twist
effects), $\propto \Lambda_{\rm QCD}^2/Q^2$ with the non-perturbative
scale  
$\Lambda_{\rm QCD}\sim 1/$fm \cite{Ellis:1982cd,Qiu:1988dn}, 
the target mass corrections (TMC), $\propto x_B^2 m_N^2/Q^2$ 
with nucleon mass $m_N$ \cite{Schienbein:2007gr}, 
and possibly, final-state jet mass corrections (JMC), 
$\propto m_j^2/Q^2$ \cite{Collins:2007ph}.
These corrections become larger and larger as data approach the
kinematic limit $x_B=1$. 
In this paper, we examine the uncertainties in extracting PDFs at low
$Q^2$ and large $x_B$ caused by the target mass and jet mass
corrections.

At the leading power, the perturbative QCD factorization treatment 
of DIS cross sections neglects all $1/Q^2$-type power corrections.
However, TMC play a somewhat special role.
Since the mass of the target is a non-perturbative quantity, the
partonic dynamics of short-distance factors in the QCD factorization
formalism should not depend on it.
Therefore, for any hadronic cross section that can be factorized 
in perturbative QCD, the effect of TMC should be implicitly included  
in the definition of the non-perturbative hadron matrix elements, and 
explicitly accounted for in the kinematic variables of the
observables. In this sense, TMC are mostly of kinematic origin.  
At large $x_B$ and low $Q^2$, the $x_B^2 m_N^2/Q^2$-type
TMC can be an important part of the measured cross sections, and 
should be identified and removed before we extract the leading 
power PDFs at large $x$.  

Following the pioneering work by Georgi and Politzer (GP) in 
as early as 1976 \cite{Georgi:1976ve}, many papers have been 
written on TMC, in particular, for lepton-hadron DIS.
A recent review by Schienbein {\it et al.} provides a 
nice summary of this effort \cite{Schienbein:2007gr}.
Most existing calculations use the technique
of operator product expansion (OPE)
to resum $m_N^2/Q^2$ corrections to the structure function moments. 
A strong debate 
has been centered on the inversion of the moment formula
\cite{DeRujula:1976tz,Gross:1976xt,Johnson:1979ty,Bitar:1978cj,Steffens:2006ds}.  
If we keep the target mass in the DIS kinematics,
the Bjorken scaling variable $x_B$ for the DIS cross sections 
or structure functions needs to be replaced 
by the Nachtmann variable \cite{Nachtmann:1973mr},
$\xi=2x_B/(1+\sqrt{1+4x_B^2 m_N^2/Q^2}) \rightarrow x_B$ as
$m_N^2/Q^2\rightarrow 0$. 
If the target mass cannot be neglected at low $Q^2$,
the Nachtmann variable $\xi$ is less than 1 even at $x_B=1$.
Only if one ignores the $x_B=1$ kinematic threshold, and allows $\xi$ to
run up to 1, does the inverse Mellin transformation of the structure
function moments give back the structure functions in $x_B$ space
\cite{Johnson:1979ty,Bitar:1978cj}.
As a consequence, the inverted structure functions are finite in the  
unphysical $x_B>1$ region.  
The unphysical region has been argued to disappear with the inclusion of
power-suppressed higher-twist terms in the computation
\cite{DeRujula:1976tz}. Alternatively, many prescriptions have been
suggested to fix the moments inversion problem, or to
phenomenologically eliminate the unphysical region, see
\cite{Johnson:1979ty,Bitar:1978cj,Steffens:2006ds}. None of these
prescriptions is entirely satisfactory or unique. 

To completely avoid the ambiguities in connection with the structure
functions moments and their inversion, it is natural to  
investigate the TMC in the momentum space without using the OPE 
and taking the moments.  
This is most easily done in the context of the
field theoretic pQCD parton model, as pioneered by Ellis, 
Furmanski, and Petronzio in Ref.~\cite{Ellis:1982cd}. Recently, 
Kretzer and Reno applied and compared both approaches in the case 
of neutrino initiated DIS experiments
\cite{Kretzer:2002fr,Kretzer:2003iu}.  
In this paper, we revisit the TMC in DIS in terms of the perturbative 
QCD collinear factorization approach in momentum space and express 
the long distance physics of structure functions and the leading 
target mass correction in terms of PDFs that share the same 
partonic operators with the PDFs of zero hadron mass. 
In our approach, the momentum space structure functions have 
no unphysical region. 
Moreover, our approach can be generalized to semi-inclusive DIS and
hadronic collisions, where the OPE is not applicable.

In the collinear factorization approach at the leading power in $1/Q^2$, the 
short-distance factors are perturbatively calculated with massless
final-state light partons.  As recently pointed by Collins, Rogers and
Stasto in Ref.~\cite{Collins:2007ph}, the outgoing parton lines should
acquire jet subgraphs/functions to have correct kinematics.  
The invariant mass in the jet subgraph leads to the before mentioned 
$m_j^2/Q^2$-type JMC, which are particularly sensitive to the 
large-$x_B$ kinematics and the extraction of large-$x$ PDFs.  
In this paper, we discuss the role of the jet functions in modifying
the DIS kinematics in the collinear factorization approach.  
We neglect the soft interactions 
between the beam jet and the final-state jet functions, and 
present a collinear factorization formalism for calculating 
DIS structure functions with a non trivial jet function. 
Based on a toy-model estimate, we argue 
that the JMC has a significant effect on the extraction of 
PDFs when $x\gtrsim 0.6$. The connection of the jet function with
lattice QCD computations of the non-perturbative quark propagator is
also discussed. 

The rest of our paper is organized as follows.  In Sec.~\ref{sec:TMC},
we drive the TMC in terms of QCD collinear factorization in momentum 
space.  We explicitly demonstrate that our result has no unphysical 
region for the DIS structure functions.  We compare our result with
TMC predicted by other approaches.  In Sec.~\ref{sec:JMC}, we discuss
the JMC. Finally, we present our summary and thoughts on 
future extensions in Sec.~\ref{sec:conclusions}. In the main text  we
limit the discussion to light partons and the transverse and longitudinal
structure functions. In the appendices, we generalize our formulae.


\section{target mass corrections}
\label{sec:TMC}

\begin{figure}[tb]
  \vspace*{0cm}
  \centerline{
  \includegraphics
    [width=\linewidth]
    {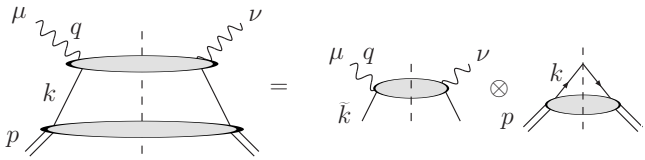}
  }
  \caption[]{
    Collinear factorization of the hadronic tensor in the impulse
    approximation. The top blob represents the interaction of a
    virtual boson with a parton computed in pQCD at any order in
    $\alpha_s$. 
  }
 \label{fig:hadronictensor}
 \label{fig:DISfactorization}
\end{figure}

The DIS cross section is determined by the hadronic tensor
\begin{align}
  W^{\mu\nu}(p,q) & = \frac{1}{8\pi} \int d^4z\, e^{-iq\cdot z}
    \vev{p|J^{\dagger\mu}(z)J^\nu(0)|p} \ ,
 \label{eq:Wdef}
\end{align}
where $p$ is the nucleon 4-momentum, $q$ is the virtual boson 
4-momentum, $J^\mu$ is the electromagnetic or electroweak current, and
$\ket{p}$ is the hadron wave function.
In the impulse approximation the lepton-nucleon interaction proceeds
through the scattering of the virtual boson with a parton (quark or
gluon) belonging to the nucleon, and having 4-momentum $k$, see
Fig.~\ref{fig:hadronictensor}. With these 4-momenta we can build the
following useful invariants:
\begin{align}
    x_B & = \frac{-q^2}{2 p\cdot q}, \quad
    Q^2 = -q^2, \quad 
    m_N^2 = p^2, \quad 
    x_f = \frac{-q^2}{2 k\cdot q } \ .
 \label{eq:invariants}
\end{align}
The first 3 invariants, namely, the Bjorken variable $x_B$, the rest
mass $m_N$ of the nucleon and the vector boson virtuality $Q^2$, are
experimentally measurable. We call them ``external invariants''. 
The fourth invariant, $x_f$, is the Bjorken variable for a partonic
target and is not experimentally measurable, so we call it
``internal''. 

We work in a class of frames, called collinear frames, 
defined such that $p$ and $q$ do not have transverse momentum. 
Then we can decompose $p$, $q$ and $k$ as follows.
\begin{align}
\begin{split}
  p^\mu & = p^+ \nbar^\mu 
          + \frac{m_N^2}{2 p_A^+} n^\mu \\
  q^\mu & = - \xi p^+ \nbar^\mu 
          + \frac{Q^2}{2\xi p^+} n^\mu \\
  k^\mu & = x p^+ \nbar^\mu 
            + \frac{k^2 + k_T^2}{2 x p^+} n^\mu 
            + \vec k_\perp^{\,\mu} \ .
 \label{eq:kinematics} 
\end{split}
\end{align}
The light-cone vectors $n^\mu$ and $\nbar^\mu$ satisfy
\begin{align}
  n^2 = \nbar^2 = 0 \qquad n\cdot\nbar=1 \ ,
\end{align}
and define the light-cone plus and minus directions, respectively. 
The plus- and minus-components of a 4-vector $a$ are defined by
\begin{align}
  a^+ = a\cdot n \qquad a^-=a\cdot\nbar .
\end{align}
If we choose $\nbar = (1/\sqrt{2},\vec0_\perp,1/\sqrt{2})$ and  
$n = (1/\sqrt{2},\vec0_\perp,-1/\sqrt{2})$, we obtain 
$a^\pm = (a_0 \pm a_3)/\sqrt{2}$. The transverse parton momentum $k_T$
satisfies $k_T\cdot n = k_T \cdot \nbar = 0$. 
The nucleon plus-momentum, $p^+$, can be interpreted as a 
parameter for boosts along the $z$-axis, connecting the target rest
frame to the hadron infinite-momentum frame.
The parton fractional light-cone
momentum with respect to the nucleon is defined as
\begin{align}
  x = k^+ / p^+ \ ,
\end{align}
and is an internal variable. The virtual boson fractional momentum
\begin{align}
  \xi = - \frac{q^+}{p^+} 
      = \frac{2 x_B} {1+\sqrt{1+4 x_B^2 m_N^2 / Q^2}}   
\end{align}
is an external variable, 
and coincides with the Nachtmann variable \cite{Nachtmann:1973mr}. 
Note that in the Bjorken limit
($Q^2 \ra\infty$ at fixed $x_B$) $\xi \ra x_B$ and we recover the
standard kinematics in the massless target approximation. 
In this paper, we will consider light quarks $u,d,s$ only and set
$m^2_{u,d,s}=0$. In Appendix~\ref{app:strfns} we will extend our
results to heavy quarks.

Collinear factorization for the hadronic tensor can be obtained by
expanding the parton momentum $k$ 
in Fig.~\ref{fig:DISfactorization} around its positive light-cone component, 
\begin{align}
  \widetilde k^\mu = x p^+ \nbar^\mu \ .
\end{align}
Correspondingly, we can define the collinear invariant
\begin{align}
  \widetilde x_f = \frac{-q^2}{2\widetilde k\cdot q} = \frac{\xi}{x} \ .
 \label{eq:tildexf}
\end{align}
According to the QCD factorization theorem \cite{Collins:1989gx},  
the nucleon hadronic tensor can then be factorized as follows: 
\begin{align}
\begin{split}
  & W^{\mu\nu}(p,q) \\ 
  & \qquad = \sum_f \int \frac{dx}{x} \, 
    \HH_f^{\mu\nu}(\widetilde k,q) \,  \varphi_{f/N}(x,Q^2,m_N^2) \\
  & \qquad  + O(\Lambda^2/Q^2)
 \label{eq:pQCDfactnolims}
\end{split}
\end{align}
where $\HH_f^{\mu\nu}$ is the short-distance partonic 
tensor for scattering on a
parton of flavor $f$, and $\varphi_{f/N}$ is the leading twist parton
distribution function for a parton of flavor $f$ inside a
nucleon $N$, see
Fig.~\ref{fig:DISfactorization}. For example,
the quark distribution at leading order in $\alpha_s$ is defined as 
\begin{align}
  \varphi_q(x,Q^2,m_N^2) = \int \frac{dz^-}{2 \pi} e^{-ix p^+ z^-}
    \vev{p|\overline\psi(z^- n)\,\frac{\gamma^+}{2}\,\psi(0)|p} \ .
 \label{eq:quarkPDFatLO}
\end{align}
A proper gauge link between the two fermion field operators is
required to have a gauge-invariant parton distribution, but drops out
if one chooses the light-cone gauge $n\cdot A = 0$, where $A$ is the
gluon field.
Higher orders in the Taylor expansion are suppressed by powers of
$\Lambda^2/Q^2$, with $\Lambda$ a hadronic scale, and contribute to
restore gauge invariance in higher twist terms \cite{Qiu:1988dn}. We will
discuss in detail how to obtain such a factorized form in
Section~\ref{sec:JMC}. 
In Eq.~\eqref{eq:pQCDfactnolims}, the partonic tensor $\HH^{\mu\nu}$ can be
computed perturbatively to any order in $\alpha_s$, and can depend on the
nucleon mass only kinematically through the invariant $\widetilde
x_f$. Dynamical target mass corrections can enter only through the
proton wave function $\ket{p}$, whence the explicit dependence of
$\varphi$ on $m_N^2$ in
Eqs.~\eqref{eq:pQCDfactnolims}-\eqref{eq:quarkPDFatLO}.
From now on we will suppress such dependence for ease of notation. 
For higher twist terms, the situation
is more complicated, because the equations of motion may induce
dynamical correlations between lower- and higher-twist terms
\cite{Ellis:1982cd}, but we will not discuss this issue here. 

\begin{figure}[tb]
  \vspace*{0cm}
  \centerline{
  \includegraphics
    [width=0.6\linewidth]
    {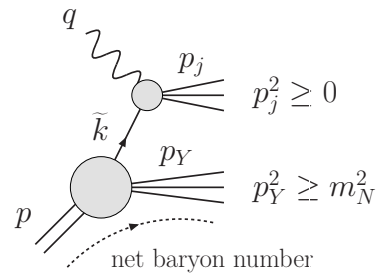}
  }
  \caption[]{
    DIS in the impulse approximation, for the special case of an
    internal on-shell light parton, $k^2 = 0$, relevant to collinear
    factorization. The current jet has momentum
    $p_j$ and the target jet has momentum $p_Y$. The net baryon number
    is only shown to flow in the target jet (lower part of the graph).
  }
 \label{fig:DISamplitude}
\end{figure}

Structure functions are obtained by suitable projections of the
tensors in Eq.~\eqref{eq:pQCDfactnolims}, see Appendix~\ref{app:strfns}. 
In this paper, we choose the helicity basis to perform the projection
of the $W^{\mu\nu}$ and $\HH^{\mu\nu}$ tensors. The transverse and
longitudinal structure functions read
\begin{align}
\begin{split}
  F_{T,L}(x_B,Q^2,m_N^2)
    = \sum_f \int \frac{dx}{x} h_{f|T,L}(\widetilde x_f,Q^2)  
    \varphi_f(x,Q^2) \ .
 \label{eq:FTL_naive}
\end{split}
\end{align}
The advantage of the helicity basis is that 
in the right hand side there are no kinematic prefactors, 
which would appear when considering the $F_{1,2}$
structure functions, as discussed in Ref.~\cite{Aivazis:1993kh} and reviewed in
Appendix~\ref{app:strfns}. 

\begin{figure*}[tb]
  \vspace*{0cm}
  \centerline{
  \includegraphics
    [width=0.45\linewidth]
    {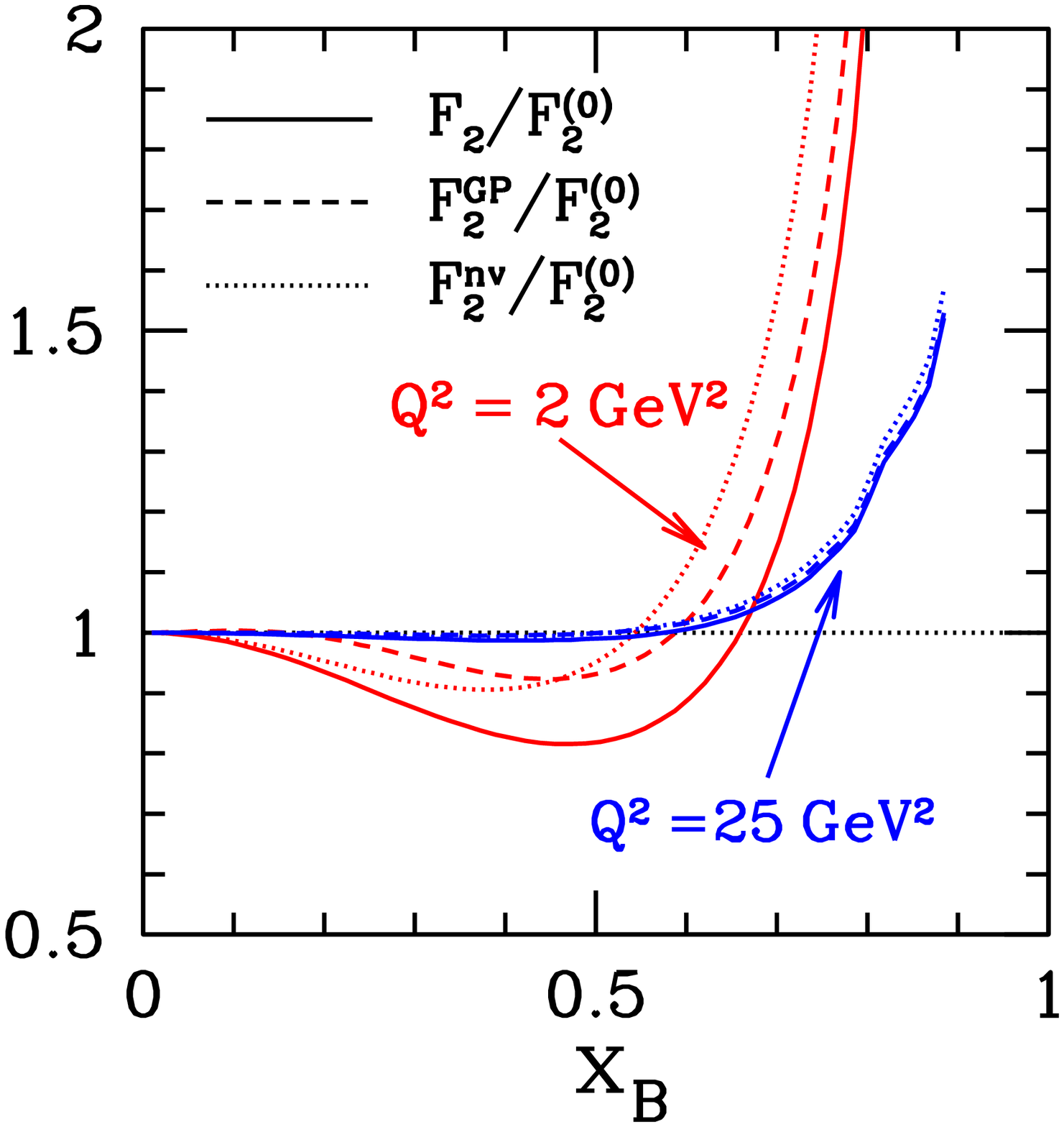}
  \hspace*{0.5cm}
  \includegraphics
    [width=0.45\linewidth]
    {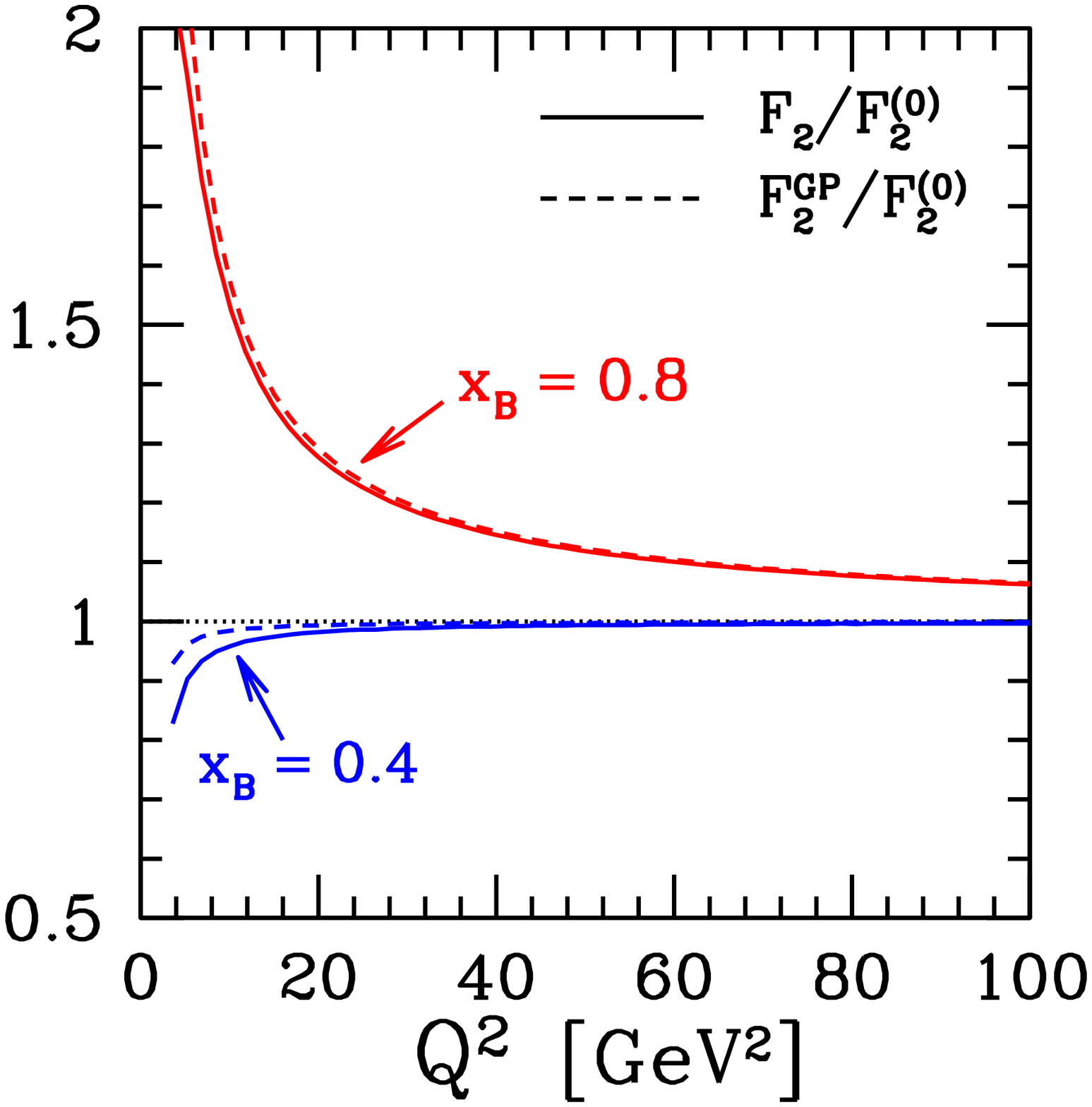}
  }
  \caption[]{
    Comparison of prescriptions for NLO target mass corrections to the
    $F_2$ structure function. The ratio $F_2/F_2^{(0)}$ is plotted as a
    function of $x_B$ and $Q^2$. The structure functions have
    been computed using MRST2002 parton distributions \cite{Martin:2002aw}. 
  }
 \label{fig:F2_TMC}
\end{figure*}

Applying the factorized Eq.~\eqref{eq:pQCDfactnolims} 
without paying attention to the kinematic
limits on $x$, which have been understood in
Eq.~\eqref{eq:pQCDfactnolims}, and using Eq.~\eqref{eq:tildexf}, one
would obtain 
what we call the ``na\"ive'' TMC in collinear factorization:  
\begin{align}
  F^\text{nv}_{T,L}(x_B,Q^2,m_N^2) =  F_{T,L}^{(0)}(\xi,Q^2)  \ ,
  \label{eq:FFnv}
\end{align}
where $F_{T,L}^{(0)}$ are the structure functions as they would be defined
and computed in the massless nucleon limit by setting $m_N^2=0$ from
the beginning. Indeed, the partonic structure functions $h_{f|T,L}$
are independent of the hadron target, and are defined in the same way 
for the massive and massless nucleon cases. 
As a consequence of the fact that $F_{T,L}^{(0)}(y,Q^2)$ has
support over $0 < y \leq 1$, the target mass corrected
$F^\text{nv}_{T,L}$ can be different from zero in the kinematically
forbidden region $1 < x_B \leq 1/(1-m_N^2/Q^2)$. The appearance of
such an unphysical region is also a feature of the OPE approach 
\cite{Georgi:1976ve,DeRujula:1976tz}, as discussed in the introduction.
Eq.~\eqref{eq:FFnv} has been introduced in Ref.~\cite{Aivazis:1993kh}
and compared to the OPE approach in
Refs.~\cite{Kretzer:2002fr,Kretzer:2003iu}. 

\begin{figure*}[tb]
  \vspace*{0cm}
  \centerline{
  \includegraphics
    [width=0.45\linewidth]
    {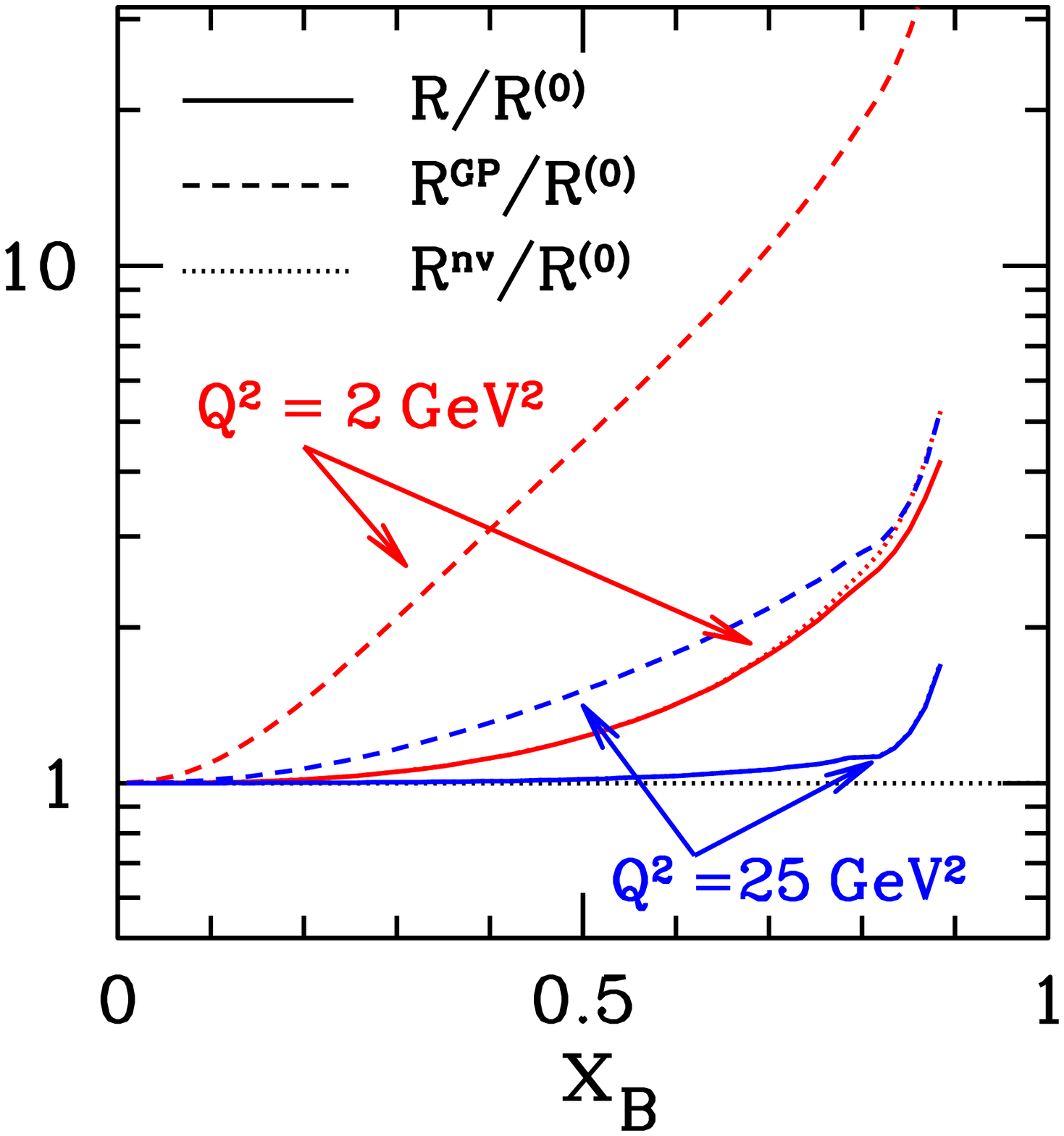}
  \hspace*{0.5cm}
  \includegraphics
    [width=0.45\linewidth]
    {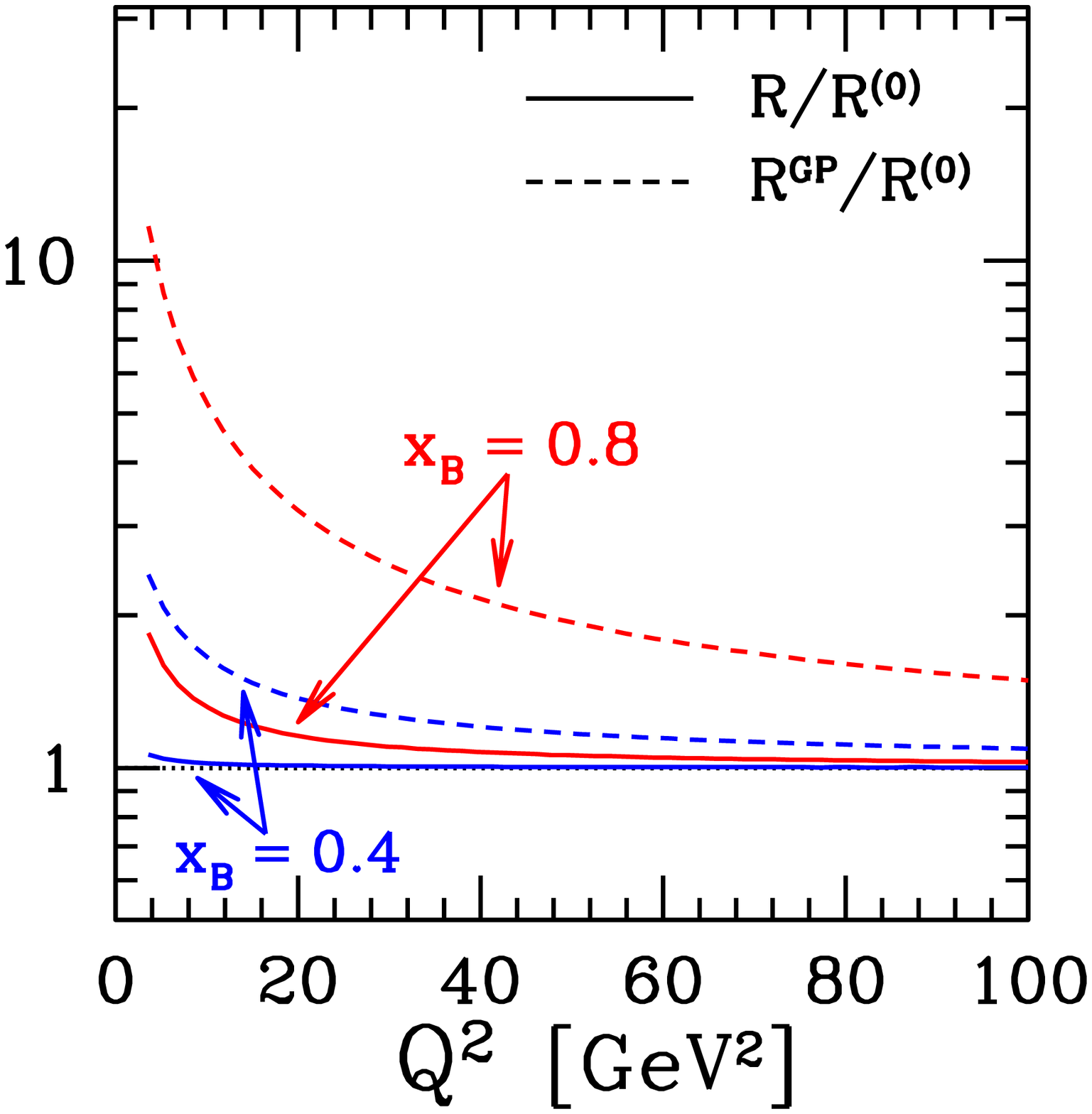}
  }
  \caption[]{
    Comparison of prescriptions for NLO target mass corrections to the
    ratio of the longitudinal and transverse cross sections,
    $R=\sigma_L/\sigma_T= F_L/F_1$. The ratio $R/R^{(0)}$ is plotted 
    as a function of $x_B$ and $Q^2$. The structure functions
    have been computed using MRST2002 parton distributions
    \cite{Martin:2002aw}.  
  }
 \label{fig:R_TMC}
\end{figure*}

In fact, a closer examination of the handbag diagram kinematics 
reveals that
there is no unphysical region. Let us consider the handbag diagram in
the right hand side of  Fig.~\ref{fig:DISfactorization}, and 
limit the discussion to on-shell light quarks or gluons,
$k^2=0$, in both the initial and final states.
The general case of off-shell partons, including heavy quark
production is discussed in Appendix~\ref{app:kinematics}.
Because of baryon number conservation, the net baryon number must 
flow either into the target jet or into the current jet. We shall
separately examine these two cases. 
If the net baryon number flows into the target jet (bottom part of
Fig.~\ref{fig:DISamplitude}), 
the jet invariant masses satisfy $m_j^2 = p_j^2 \geq m_f^2$ and 
$p_Y^2 \geq m_N^2$.
Let us consider the invariant momentum square of the process,
$s = (p+q)^2 = (p_j + p_Y)^2$. Since the 2 jets are
made of on-shell particles, $p_j\cdot p_Y \geq0$. Hence,
$s \geq m_j^2 + m_N^2$. In summary, the current 
jet mass must satisfy
\begin{align}
  0 \leq m_j^2 \leq s - m_N^2  \ . 
 \label{eq:mj_bdown}
\end{align}
Since $s-m_N^2 = (1/x_B -1)Q^2$, Eq.~\eqref{eq:mj_bdown} 
guarantees that the handbag diagram is non-zero only when
$x_B \leq 1$, as it must be
on general grounds because of baryon number conservation, irrespective
of the model used to compute the process. 
On the other hand, if the net baryon
number flows into the current jet (top part of
Fig.~\ref{fig:DISamplitude}). The invariant jet masses satisfy 
$m_j^2 \geq m_N^2$ and $p_Y^2 \geq 0$, so that
\begin{align}
  m_N^2 \leq m_j^2 \leq s \ ,
 \label{eq:mj_bup}
\end{align}
which again guarantees that the handbag diagram respects the 
$x_B \leq 1$ limit. 
Within the collinear factorization approach, 
the momentum of the active quark entering the short-distance 
hard part that generates the current jet is approximated to be on mass
shell, $\tilde{k}^2=0 \ll m_N^2$.  
That is, the baryon number is very likely to flow into 
the target jet for the factorized contribution to the 
DIS cross section, and 
Eq.~\eqref{eq:mj_bdown} gives the relevant limits on $m_j^2$.
Using $m_j^2 = (\tilde k+q)^2 = (1/\widetilde x_f-1)Q^2$ and $\widetilde
x_f=\xi/x$ in Eq.~\eqref{eq:mj_bdown}, we obtain 
\begin{align}
  x_B \leq \widetilde x_f \leq 1 \ ,
 \label{eq:xflimits}
\end{align}
which implies the following limits on the $dx$ integration in
Eq.~\eqref{eq:FTL_naive}: 
\begin{align}
  \xi \leq x \leq \frac{\xi}{x_B} 
 \label{eq:xlimits}
\end{align}
Eqs.~\eqref{eq:xflimits}-\eqref{eq:xlimits}
explicitly guarantee $F_{T,L} = 0$ if $x_B > 1$, so that there is no
unphysical region for target mass corrected structure functions:
\begin{align}
\begin{split}
 & F_{T,L}(x_B,Q^2,m_N^2) \\
 & \qquad = \int_\xi^{{\xi}/{x_B}} \frac{dx}{x} h_{f|T,L}(\widetilde x_f,Q^2)  
    \varphi_f(x,Q^2) \ .
 \label{eq:FTL_TMC}
\end{split}
\end{align}
Eq.~\eqref{eq:FTL_TMC} is our formula for calculating DIS structure 
functions with the TMC.  As expected, it has the hadron mass dependence 
explicitly in the integration limits caused by the DIS kinematics and 
implicitly from the hadron states in the definition of the PDFs.  
The na\"ive structure functions \eqref{eq:FFnv} are obtained when
considering $x\leq 
1$ as upper integration limit in Eq.~\eqref{eq:FTL_TMC}. This limit is
a general and process-independent consequence of the 
definition of a parton distribution in the field theoretic parton
model \cite{Jaffe:1983hp}, but in DIS it is weaker than $x \leq
\xi/x_B$, which is induced by 4-momentum and baryon number
conservation.
In the massless target limit, $m_N^2/Q^2 \ra 0$, 
the constraint \eqref{eq:xlimits} reduces to $x_B \leq x \leq 1$,
and we recover the massless structure functions as we should expect:
\begin{align}
  F_{T,L}(x_B,Q^2,m_N^2) \xrightarrow[m_N^2/Q^2\ra 0]{}  
    F_{T,L}^{(0)}(x_B,Q^2)  \ .
\end{align}

In Fig.~\ref{fig:F2_TMC}, we plot the ratio of the TMC corrected
$F_2$ to the massless $F_2^{(0)}$, with TMC computed using the analog
of Eq.~\eqref{eq:FTL_TMC}, see Appendix~\ref{app:colfactsfn}, 
the naive prescription \eqref{eq:FFnv}, and
the Georgi-Politzer prescription. The corrections are in general quite
large at $Q^2=2$, but still non negligible at the generally considered
``safe'' scale $Q^2=25$ GeV$^2$. 
From the right panel of the figure, one can estimate how large
$Q^2$ should be to safely neglect TMC. At $x_B\lesssim
0.5$ the TMC are smaller than 5\% if $Q^2\gtrsim 10$ GeV$^2$. However,
at larger $x_B$, one may need to go to $Q^2 \gtrsim 100$
GeV$^2$ for TMC to become small. Note also the difference between
$F_2$ and $F_2^\text{nv}$, which is smaller than 30-40\% at $Q^2=2$:
it gives the size of the contribution of the unphysical region
$\xi/x_B < x \leq 1$, which has to be subtracted from the na\"ive
structure function.

The difference between TMC of $F_2$ (and similarly of $F_{1,T}$)
in collinear factorization and in the Georgi-Politzer formalism is
smaller than 15-20\% at the lowest scale, and rapidly disappears at
larger scales. So one
is tempted to brush aside the question of what formalism is correct,
if willing to accept this level of uncertainty. However, the
situation completely changes when considering $F_L$, or the ratio $R$ 
of the longitudinal to transverse cross section, 
\begin{align}
\begin{split}
  R = \frac{\sigma_L}{\sigma_T} = \frac{F_L}{F_1}  \ ,
  \qquad
  R^{(0)} = \frac{F^{(0)}_L}{F^{(0)}_1} \ ,
\end{split}
\end{align}
whose TMC/massless ratios are plotted in Fig.~\ref{fig:R_TMC}.
(Note a factor of $2x_B$ with respect to other common conventions, see
Appendix~\ref{app:invsfn}.) The TMC of $R$ 
are much larger than for $F_2$. Most importantly, the
difference between the collinear factorization and Georgi-Politzer TMC
is huge, up to a factor 10 (5)  at $Q^2$=2 (25) GeV$^2$! Therefore,
one has to decide which formalism to use. This is especially important
for a fit of the gluon PDF, to which $F_L$ is sensitive.

\begin{figure}[tb]
  \vspace*{0cm}
  \centerline{
  \includegraphics
    [width=\linewidth]
    {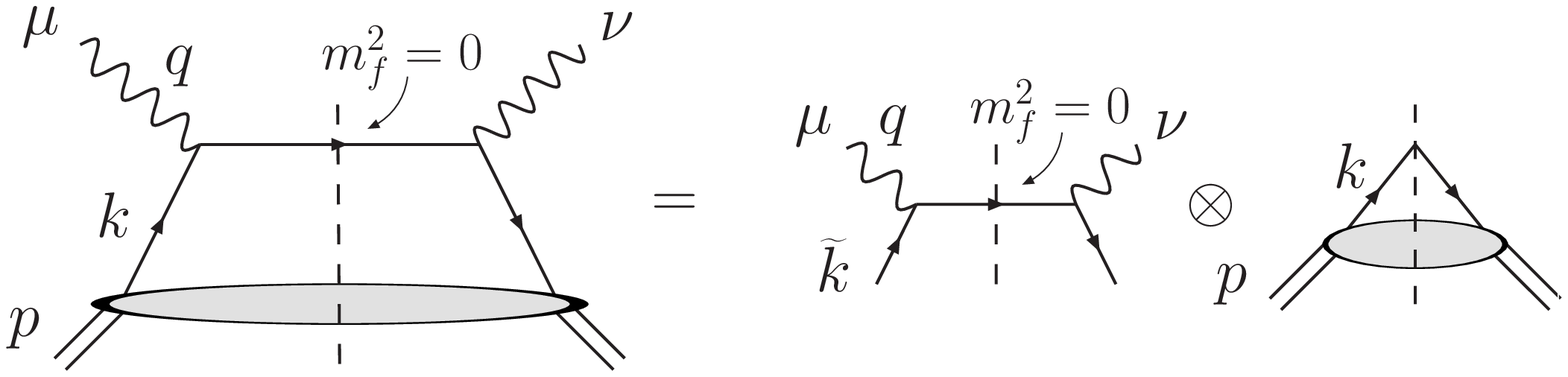}
  }
  \caption[]{
    DIS handbag diagram at leading order in $\alpha_s$. 
  }
 \label{fig:DIS_LO}
\end{figure}

\begin{figure}[tb]
  \vspace*{0cm}
  \centerline{
  \includegraphics
    [width=0.9\linewidth]
    {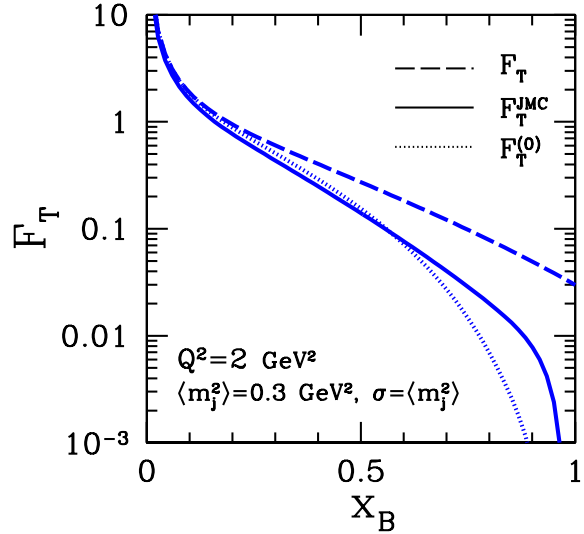}
    }
  \caption[]{
    Transverse structure function plotted as a function of $x_B$,
    with and without target and jet mass corrections, computed with
    only light quarks at lowest order in $\alpha_s$ using MRST2001LO
    parton distributions \cite{Martin:2002dr}. The dotted line is the
    massless structure function. The dashed line includes only TMC,
    and corresponds to $Z=1$ in Eq.~\eqref{eq:JMCnosoft}. The solid line
    corresponds to JMC coming only from the continuum part $\rho$ of
    the jet spectral function, $Z=0$ in Eq.~\eqref{eq:JMCnosoft}. JMC
    are computed using a log-normal spectral function with
    $\vev{m_j^2} = 0.3$ GeV$^2$ and standard deviation
    $\sigma_{m_j^2}=\vev{m_j^2}$. 
  }
 \label{fig:FTvsFT0}
\end{figure}

It is also important to note that our formula for TMC 
in Eq.~\eqref{eq:FTL_TMC} explicitly eliminates the kinematically 
forbidden region $1 < x_B \leq 1/(1-m_N^2/Q^2)$ because of the 
integration limits on the parton momentum fraction $x$.  
As $x_B\rightarrow 1$, structure functions calculated by 
using Eq.~\eqref{eq:FTL_TMC} approach to zero, the kinematic limit, 
smoothly, except for the lowest order contribution, whose partonic 
structure functions are derived from the tree level handbag 
diagram in Fig.~\ref{fig:DIS_LO}. 
Indeed, by explicit computation at tree level in the approximation of
massless quarks, we obtain  
\begin{align}
  h_{q|T}(\widetilde x_f,Q^2) 
    = \frac{1}{2}\, e_f^2\, \delta(\widetilde x_f - 1) 
    = \frac{1}{2}\, e_f^2\, x\, \delta(x - \xi) \ ,
 \label{eq:hTLO}
\end{align}
where $e_f$ is the electric charge of the parton $f$.  
Substituting the tree-level partonic structure function into
Eq.~\eqref{eq:FTL_TMC}, the lowest order contribution to 
the transverse structure function, 
\begin{align}
  F_T(x_B,Q^2,m_N^2) 
    = \left\{\bay{cl}
      F_T^{(0)}(\xi,Q^2) \quad & x_B \leq 1 \\
      0                       &  x_B > 1 \ ,  \\
    \eay \right.
\end{align}
remains positively finite when $x_B\rightarrow 1$ and does
not vanish as $x_B\ra1$, as shown by the dashed line 
in Fig.~\ref{fig:FTvsFT0}, 
  
\begin{figure}[tb]
  \vspace*{0cm}
  \centerline{
  \includegraphics
    [width=0.7\linewidth]
    {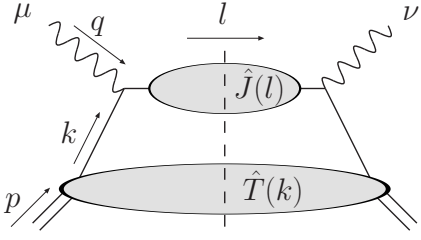}
  }
  \caption[]{
    DIS handbag diagram at leading order in $\alpha_s$ generalized to
    include a jet function $\hat J(l)$ beside the target function
    $\hat T(k)$.  
  }
 \label{fig:DIS_JET_LO}
\end{figure}

This problem exists only at the lowest order and 
arises because of the $\delta$-function behavior of the partonic
structure function \eqref{eq:hTLO} and the assumption that 
the final-state is made of a massless quark, $m_f^2=0$, 
as shown in Fig.~\ref{fig:DIS_LO}.  The $\delta$-function
bypasses the kinematic constraint from the integration limits
in Eq.~\eqref{eq:FTL_TMC} and forces $x=\xi(x_B)$, which exhibits 
the mismatch between the phase space for $x$ at the parton level
and that for $\xi(x_B)$ at the hadron level.
Under the collinear approximation the momentum fraction
$x$ for a massless parton can be as large as 1, while the plus
momentum fraction of the virtual photon $\xi(x_B)$ smaller than 1  
for a finite target mass $m_N$.
As a result, the perturbatively calculated structure functions 
do not vanish at $x_B=1$ because the PDFs are finite at
$x=\xi(x_B=1)=2/(1+\sqrt{1+4m_N^2/Q^2}) < 1$.  
As we will discuss in the next section, this explicit 
phase space mismatch at the lowest order could be improved
if the single massless quark final-state in Fig.~\ref{fig:DIS_LO}, 
which is not physical, is replaced by a jet function as shown 
in Fig.~\ref{fig:DIS_JET_LO}.

We conclude this section by stating that 
if one is performing global QCD fits of the PDFs
in the context of pQCD collinear factorization, our formalism in 
Eq.~\eqref{eq:FTL_TMC} might be the most consistent way to treat TMC, 
because it expresses the long distance physics of structure functions 
and the leading target mass correction in terms of PDFs that 
share the same partonic operators with the PDFs of zero hadron mass. 
Moreover the structure functions calculated using our formulae 
do not have the $x_B>1$ unphysical region and vanish at the 
$x_B=1$ kinematic limit except for the lowest order contribution 
that will be discussed further in next section.
The same collinear factorization formalism can be easily and
consistently extended to semi-inclusive DIS measurements and hadronic
collisions, for which the OPE formalism is not applicable, but which
are included in global QCD fits of parton distributions. Careful
analysis of kinematics and conservation laws will guarantee that no
unphysical region appears in these observables, as well. 
The obtained formulae will not merely be an approximation to the TMC
for those processes, as argued in
\cite{Kretzer:2003iu,Kretzer:2002fr}, but will give the correct answer
in the context of pQCD collinear factorization.

\section{Jet mass corrections}
\label{sec:JMC}

In this section we discuss the possibility to include 
a jet function into the lowest order contribution to have a 
more realistic kinematic constraint on the ``single quark'' 
final-state \cite{Collins:2007ph}.  
Hopefully, we can reduce the unphysical positive 
value of structure functions at $x_B=1$. As discussed in the last section,
this is caused 
by the $\delta$-function behavior of the partonic structure 
functions, the assumption that the final-state is made of 
a massless quark, $m_f^2=0$, and the mismatch between the 
phase space for $\xi(x_B)$ and $x$.   

The assumption that the leading order final-state is made of 
a massless quark, $m_f^2=0$, is clearly unphysical 
because the quark has to hadronize due to color confinement, 
so that the current jet will have an invariant mass $m_j^2$.
Then, we may heuristically set $m_f^2 =
m_j^2$ for the cut quark line, and substitute $\delta(\widetilde x_f
- 1)$ with $\delta(\widetilde x_f - 1/(1+m_j^2/Q^2))$ in
Eq.~\eqref{eq:hTLO}: 
\begin{align}
  h_{q|T}(\widetilde x_f,Q^2) \lora 
    \frac{1}{2}\, e_f^2\, x\, 
    \delta \big( x - \xi(1+\frac{m_j^2}{Q^2}) \big) \ .
 \label{eq:hTLO_jet}
\end{align}
Furthermore, we may assume that the current jet has
an invariant mass probability distribution $J_m(m_j^2)$ normalized to 1,
and accordingly smear the structure functions in \eqref{eq:FTL_TMC}:
\begin{align}
\begin{split}
 & F_T^{JMC}(x_B,Q^2,m_N^2) \\
 & \qquad = \int_0^\infty dm_j^2 J_m(m_j^2) 
   \int_\xi^{{\xi}/{x_B}} \frac{dx}{x} h_{f|T}(\widetilde x_f,Q^2)  
    \varphi_f(x,Q^2) \\
 & \qquad = \int_0^{\frac{1-x_B}{x_B}Q^2} dm_j^2 J_m(m_j^2) 
   F_T^{(0)}\big(\xi(1+m_j^2/Q^2),Q^2\big) \ .
 \label{eq:FTL_TMC_JMC_heuristic}
\end{split}
\end{align}
If $J_m(m_j^2)$ is a sufficiently smooth function of $m_j^2$, we obtain 
\begin{align}
  F_T^{JMC}(x_B,Q^2,m_N^2) \xrightarrow[x_B\ra 1]{} 0 \ . 
\end{align}
The jet mass corrections (JMC) so introduced are of order
$O(m_j^2/Q^2)$. It is easy to see that in the limit $Q^2 \gg
\vev{m_j^2}$, the massless $F_T^{(0)}$ decouples from the integration
over the jet mass, and we recover the structure function with TMC:
\begin{align}
   F_T^{JMC}(x_B,Q^2,m_N^2) \xrightarrow[Q^2 \gg \vev{m_J^2}]{}
    F_T(x_B,Q^2,m_N^2) \ .
\end{align}

In the following, we will discuss how to put this Ansatz on a more
firm theoretical basis.

\subsection{Collinear factorization with a jet function}
\label{sec:jetfact}

We aim at including in the DIS handbag at leading order in
$\alpha_s$ a suitable jet function to take into account the
invariant mass of the jet produced by the hadronization of the struck
quark, see Fig.~\ref{fig:DIS_JET_LO}. 
Note that in computing the DIS cross section with the handbag diagram
of Fig.~\ref{fig:DIS_JET_LO}, we are making several assumptions. 
First, we are assuming that it makes sense to separate the final state
into a current jet and a target jet, respectively the top and the
bottom blob. Because of color confinement, this separation can only
make sense as an approximation, and is justified for inclusive and
semi-inclusive cross sections 
if the rapidity separation between the 2 jets is large 
enough. This is in general the case at asymptotically large
$Q^2$. However, at finite $Q^2$, the rapidity difference between the 2
jets tends to 0 as $x_B\ra 1$, and the struck quark may participate
in the hadronization process together with the unstruck target
partons. Thus, we need to take care in estimating the range in
$x_B$ in which the handbag diagram is a meaningful approximation to
the DIS process. The second assumption we make, intimately related
with the first one, is that color neutralization of the current jet
happens via the exchange of soft momenta, which we can neglect when
discussing 4-momentum conservation. 

In order to obtain a
collinear factorization formula, we will closely follow the
procedure of Ellis, Furmanski and Petronzio \cite{Ellis:1982cd}.
The hadronic tensor is
\begin{align}
  W^{\mu\nu}(p,q) = \frac{e_f^2}{8\pi} 
    \int \frac{d^4k}{(2\pi)^4} 
    \Tr \big[\hat T(k) \gamma^\nu \hat J(l) \gamma^\mu \big] \,
    {\mathbb K}(k,p,q) \ .
\end{align}
where we considered only 1 flavor for simplicity. The sum over quark
flavors will be restored at the end of the computation.
We use a hat to denote a matrix in Dirac space. The trace over color
indexes can be easily factorized and included in the target function
\cite{Ellis:1982cd}. The remaining trace is over Dirac indexes.
The target function $\hat T$ is defined as
\begin{align}
\begin{split}
  \big[\hat T(k)\big]_{ij} 
    & = \sum_Y \delta^{(4)}\big(p-k-\sum_{i\in Y}p_i\big)
      \big|\vev{p|k,Y}\big|^2  \\
    & = \int d^4z e^{iz\cdot k} \vev{p|\overline\psi_j(z)\psi_i(0)|p} \ ,
\end{split}
\end{align}
where 
$\bra{k,Y} = \bra{k}\bra{Y}$, 
$\bra{Y}$ are all possible final states originating from the
target fragmentation, 
and $\bra{k}$ is a parton state of momentum $k$. 
Analogously, the jet function $\hat J$ is the 
non-perturbative quark propagator: 
\begin{align}
\begin{split}
   \big[\hat J(l)\big]_{ij} 
     & = \sum_Y \delta^{(4)}\big(l-\sum_{i\in Y} p_i\big)
       \big|\vev{l|Y}\big|^2  \\
     & = \int d^4z e^{iz\cdot l} \vev{0|\overline\psi_j(z)\psi_i(0)|0}
     \ , 
\end{split}
\end{align}
and $\bra{l}$ is a quark state of momentum $l$. 
The jet momentum is constrained by momentum conservation to $l=k+q$,
but it is useful to keep it explicit in our formulae. The function
$\mathbb K$ is included to impose the kinematic constraints, 
the non-trivial one being 
$x_f^\text{min} \leq x_f \leq x_f\text{min}$, see
Appendix~\ref{app:kinematics}:
\begin{align}
\begin{split}
  \mathbb K(k,p,q) & = \theta(k^++q^+) \theta(k^-+q^-) \\ 
  & \times \theta(p^+ - k^+) \theta(p^- - k^-) \\
  & \times \theta(x_f-x_B) \theta(1-x_f) \ ,
\end{split}
\end{align}
where, for light quarks, 
\begin{align}
\begin{split}
  x_f^\text{min} & = \frac{x_B}{1-x_B k^2/Q^2} \\
  x_f^\text{max} & = \frac{1}{1-k^2/Q^2} \ .
\end{split}
\end{align}
To obtain the leading power contribution, we expand
$\hat T(k)$ in terms of Dirac matrices and neglect terms that 
depend on the vector defining the direction of the gauge link
in the PDFs, which are suppressed by powers of $1/Q^2$ 
\cite{Qiu:1988dn}, 
\begin{align}
\begin{split}
  \hat T(k) & = \tau_1(k) \hat {\mathbb I} + \tau_2(k) \ksl
    + \tau_3(k) \gamma_5 + \tau_4(k) \ksl \gamma_5 \, ,
\end{split}
\end{align}
and, analogously, we expand $\hat J$ as :
\begin{align}
\begin{split}
  \hat J(l) & = j_1(l) \hat {\mathbb I} + j_2(l) \lsl
    + j_3(l) \gamma_5 + j_4(l) \lsl \gamma_5 \ .\\
\end{split}
\end{align}
For massless quarks, $\tau_1 = 0$. The terms $\tau_{3,4}$, which are
proportional to $\gamma_5$ cancel when computing unpolarized cross
sections. In pure QCD, $j_{3,4}=0$ because of parity invariance, and
$j_1$ only enters in traces with an odd number of $\gamma$ matrices,
hence does not contribute. We are left with the terms proportional to
$\tau_2$ and $j_2$. The dominance of the $k^+$ and $l^-$ components of
$k$ and $l$ in the Breit frame suggests to define
\begin{align}
&\begin{split}
  \tau_2(k) & = \frac{1}{4 k^+} \Tr\big[\nsl \hat T(k)\big] \\
    & = \frac{1}{4 k^+} \int d^4z e^{iz\cdot k}
    \vev{p|\overline\psi(z)\gamma^+\psi(0)|p} \, \\
 \label{eq:tau2def}
\end{split} \\
&\begin{split}
  j_2(l) & = \frac{1}{4 l^-} \Tr\big[\nbsl \hat J(l)\big] \\
    & = \frac{1}{4 l^-} \int d^4z e^{iz\cdot l}
    \vev{0|\overline\psi(z)\gamma^-\psi(0)|0} \ .
  \label{eq:j2}
\end{split}
\end{align}
After these manipulations, the hadronic tensor reads
\begin{align}
\begin{split}
   W^{\mu\nu}(p,q) & = \int \frac{dk^+dk^-d^2k_T}{(2\pi)^4} \\
   & \times \frac{e_f^2}{8\pi} 
      \Tr \big[\ksl \gamma^\nu \lsl \gamma^\mu \big] \,j_2(l)\,
      \tau_2(k)\, {\mathbb K}(k,p,q) \ ,
 \label{eq:W-step0}
\end{split}
\end{align}
where 
\begin{align}
  k^\mu & = x p^+ \nbar^\mu 
            + \frac{k^2+k_T^2}{2 x p^+} n^\mu 
            + k_T^{\,\mu} \\
  l^\mu & = (x-\xi) p^+ \nbar^\mu 
            + \big( \frac{k^2+k_T^2}{2 x p^+} 
              + \frac{Q^2}{2\xi p^+} \big)n^\mu 
            + k_T^{\,\mu} \ . 
\end{align}
For later use, let us also define 
\begin{align}
  \frac{1}{\pi} H_*^{\mu\nu}(k,l) = \frac{e_f^2}{8\pi} 
      \Tr \big[\ksl \gamma^\nu \lsl \gamma^\mu \big] \ .
  \label{eq:H*def}
\end{align}

Our goal is to obtain a factorized expression for the hadronic tensor
in terms of collinear parton distribution functions, see for example
Eq.~\eqref{eq:quarkPDFatLO}. For this purpose we need to let
$\int dk^-d^2k_T$ act only on $\tau_2(k)$, which defines the collinear
PDF modulo factors of 2. In doing this we will be forced to make
approximations on the momenta entering and exiting the hard scattering
vertex, viz., $k$ and $l$. In principle, one would like to avoid it and
allow approximations in the computation of the hard scattering
tensor only \cite{Collins:2007ph}. In this way, one can ensure that
the final state obeys 4-momentum conservation, and avoid potentially
large errors in region of phase-space close to the kinematic
boundaries. While in most cases this is not a problem for inclusive
cross sections, it might become very important for exclusive
observables. In our case, we want to compute the inclusive DIS cross
section at large $x_B \ra 1$ in collinear factorization: 
in order to extend the validity of our
computation as close as possible to this kinematic boundary, we
need to pay attention to the approximations we will make, keep
them at a minimum, and estimate the range of validity in $x_B$ and 
$Q^2$ of the approximations we will have to make.

The first step in the collinear factorization of the hadronic tensor 
\eqref{eq:W-step0}, is to expand $H_*^{\mu\nu}$ around the momentum of
a collinear and massless quark:
\begin{align}
  H_*^{\mu\nu}(k,l) = H_*^{\mu\nu}(\widetilde k,\widetilde l)
    + \frac{\partial H_*^{\mu\nu}}{\partial k^\alpha}
      (k^\alpha - \widetilde k^\alpha) + \ldots
\end{align}
where
\begin{align}
\begin{split}
  \widetilde k^\mu & = x p^+ \nbar^\mu \\
  \widetilde l^\mu & = \widetilde k^\mu + q^\mu \ .
\end{split}
\end{align}
The higher order terms in the expansion are suppressed as powers of
$\Lambda^2/Q^2$, where $\Lambda^2$ is a hadronic scale, and contribute
to restore gauge invariance in higher-twist diagrams \cite{Qiu:1988dn}. In
this paper, we will retain only the leading twist term of the
expansion. Note that we did not yet make any kinematic
approximation: in principle, one may sum over as many higher-twist
terms as desired.

The second step involves using the spectral representation
of $\hat J$ \cite{Weinberg:1995mt} to explicitly introduce the
invariant jet mass in the formalism:
\begin{align}
  \hat J(l) = \int_0^\infty \hspace*{-.3cm}dm_j^2\, 
    \big[J_1(m_j^2) \hat{\mathbb I} + J_2(m_j^2) \lsl \big]
    2\pi \delta(l^2-m_j^2) \theta(l^0) \ ,
  \label{eq:Jhat}
\end{align}
where the spectral functions $J_i(m_j^2)$ are positive definite and
normalized to 1:
\begin{align}
  \int_0^\infty \hspace*{-.3cm}dm_j^2\,J_i(m_j^2) = 1 \ .
  \label{eq:J2norm}
\end{align}
In particular, by substituting Eq.~(\ref{eq:Jhat}) into (\ref{eq:j2}),
we obtain
\begin{align}
  j_2(l) = \int_0^\infty \hspace*{-.3cm}dm_j^2\,J_2(m_j^2)\,
    2\pi \delta(l^2-m_j^2)\, \theta(l^0) \ ,
 \label{eq:j2-spectral}
\end{align}
so that we can interpret $m_j$ as the jet invariant mass, and $J_2(m_j^2)$
as its probability distribution. 

In the light-cone
gauge $n\cdot A=0$, the jet spectral function is related to the non
perturbative quark propagator: 
\begin{align}
\begin{split}
  & \int_0^\infty \hspace*{-.3cm}
    dm_j^2\,J_2(m_j^2)\,2\pi \delta(l^2-m_j^2)\, \theta(l^0) \\
  & \qquad  =  \frac{1}{4 l^-} \int d^4z e^{iz\cdot l}
    \Tr \big[ \gamma^- \vev{ 0 | \overline\psi(z)\psi(0)|0} \big] \ .
  \label{eq:jet-qprop}
\end{split}
\end{align}
Computations of the non-perturbative quark propagator have been
performed in lattice QCD \cite{Bowman:2005vx} and using 
Schwinger-Dyson equations, see \cite{Fischer:2006ub} for a review. 
However, there are several difficulties in extracting information
relevant to the jet spectral function from these computations: (i) the
quark-antiquark correlator appearing in \eqref{eq:jet-qprop} is 
typically computed in the Landau gauge instead of the light-cone
gauge, (ii) computations are performed in Euclidean space instead of
Minkowski space, (iii) one needs to extract the spectral
representation from the computed correlator. 
The biggest problem is that the analytic structure of the quark
propagator is not sufficiently well known to either perform the
analytic continuation back to Minkowski space or to extract its
spectral representation \cite{Fischer:2006ub,Alkofer:2003jj}. 
As a way to avoid this
problem, it would be interesting to see if it is possible to rotate
the whole handbag diagram, including its external momenta, to
Euclidean space as done in \cite{Blum:2002ii} for the computation of the
hadronic contribution to the muon anomalous magnetic moment. 
In this way one
would be able to directly use the lattice propagator in the
computation of the forward Compton amplitude. Alternatively, one may 
try to use the light-cone QCD formulation on the lattice discussed in
\cite{Grunewald:2007cy}, which exploits the Hamiltonian formulation of
QCD in order to remain in Minkowski space. A more phenomenological
approach to the spectral function will be discussed in the next 
subsection.

The third step involves our first kinematic approximation. In order to
factorize $j_2$ from the $dk^-d^2k_T$ integrations, we need to
approximate $l \ra \widetilde l$, so that 
\begin{align}
  j_2(l) \,\lora\, j_2(\widetilde l) 
    = \int_0^\infty \hspace*{-.3cm}dm_j^2\,J_2(m_j^2)\,
    2\pi \delta(\widetilde l^2-m_j^2)\, \theta(l^0) \ .
 \label{eq:approx-step3}
\end{align}
Then, the hadronic tensor reads
\begin{align}
  W^{\mu\nu}(p,q) & =  \int_0^\infty\hspace*{-.3cm}dm_j^2\,J_2(m_j^2) 
    \int dk^+\, H_*^{\mu\nu}(\widetilde k,\widetilde l)\,
    \delta({\widetilde l}^2-m_j^2) 
    \nonumber \\
  & \times 
    \int \frac{dk^-d^2k_T}{(2\pi)^4} 2\tau_2(k)\, {\mathbb K}(k,p,q)
 \label{eq:W-step3}
\end{align}
where $\theta(l^0)=\theta(k^0+q^0)$ is already included 
in the kinematic constraint function $\mathbb K(k,p,q)$.
We can expect the approximation \eqref{eq:approx-step3} 
to be reasonable in a region 
where $j_2(l)$ does not vary strongly with $l$. In terms of the
spectral representation, this requirement is satisfied if the integral
in Eq.~\eqref{eq:W-step3} is dominated by values of $m_j^2$ close to where 
the jet spectral function has a maximum. We will discuss below the
conditions on $x_B$ and $Q^2$ for which this condition is satisfied.
Note that this kinematic approximation only acts on the
$\delta$-function in Eq.~\eqref{eq:W-step3} so that $J_2(m_j^2)$ has
been left unapproximated: in this sense the approximation is the
mildest possible compatible with collinear factorization.

The fourth step involves decoupling  $\mathbb K$ and $\tau_2$ in
Eq.~\eqref{eq:W-step3}. It can be achieved by
replacing
\begin{align}
  {\mathbb K}(k,q,p) \,\lora\, {\mathbb K}(\widetilde k,q,p) 
    = \theta(\widetilde x_f-x_B) \theta(1-\widetilde x_f) \ .
\end{align}
Note that $\theta(\widetilde k^0+q^0)=\theta(p^+-\widetilde k^+)
=\theta(p^--\widetilde k^-)=1$ because of the constraints on
$\widetilde x_f$. In terms of $x$,
\begin{align}
  \xi \leq x \leq \xi/x_B \ .
\end{align}
This is a delicate step because it involves approximating the
kinematic constraints, such that the integration over $k_T$ and $k^-$
are unbounded. This clearly is not a good approximation as $x_B \ra 1$
\cite{Collins:2007ph}, in which case the struck parton carries most of the
nucleon plus-momentum, so that the minus and transverse components
cannot be large. To appreciate this, consider
\begin{align}
  s = (p+q)^2 = (p_Y + l)^2 \ ,
\end{align}
where $p_Y$ is the total four momentum of the target jet, and we
define $m_Y^2 = p_Y^2 \geq 0$, see
Fig.~\ref{fig:DISamplitude}. Using the full kinematics of
Eq.~\eqref{eq:kinematics}, we obtain
\begin{align}
  s = \frac{1-\xi}{\xi} Q^2 + (1-\xi) m_N^2 \ .
\end{align}
On the other hand, in the center-of-mass frame, $\vec p_Y = - \vec l$
and $\vec p_{Y,T} = - \vec l_T = - \vec k_T$, so that 
\begin{align}
  s & = (p_Y^0 + l^0)^2 \\
    & = \Big( \sqrt{m_Y^2 + k_T^2 + (p_Y^3)^2}
    + \sqrt{m_j^2 + k_T^2 + (l^3)^2} \,\Big)^2 > 4k_T^2 
 \nonumber
\end{align}
Combining these 2 results, we obtain
\begin{align}
  k_T^2 < \frac{1-\xi}{4\xi} Q^2 \big( 1 + \xi\frac{m_N^2}{Q^2} \big)
    \ .
  \label{eq:ktbound}
\end{align}
As $x_B\ra 1$, $\xi \ra \xi_{th} \lesssim 1$ so that the $(1-\xi)$
factor tends to close the available $k_T$ phase space. 
In Section~\ref{sec:numest}, we will discuss
in which region of $x_B$ and $Q^2$ we may in fact neglect this
bound. Using the definition \eqref{eq:tau2def} of $\tau_2$,
the hadronic tensor reads
\begin{align}
  W^{\mu\nu}(p,q) & =  \int_0^\infty\hspace*{-.3cm}dm_j^2\,J_2(m_j^2) 
 \label{eq:W-step4} \\
  & \times \int_\xi^{\xi/x_B} \frac{dx}{x}  
    H_*^{\mu\nu}(\widetilde k,\widetilde l)\,
    \delta({\widetilde l}^2-m_j^2) \varphi_q(x,Q^2) \ ,
  \nonumber
\end{align}
where the quark PDF $\varphi_q$ is defined as in
Eq.~\eqref{eq:quarkPDFatLO}.

As a last step, we define an on-shell and massless jet momentum 
for the partonic tensor, 
\begin{align}
  \widehat l^\mu = \widetilde l^- n^\mu = \frac{Q^2}{2\xi p^+} n^\mu
  \, 
\end{align}
and replace 
\begin{align}
  H_*^{\mu\nu}(\widetilde k,\widetilde l) \,\lora\, 
    H_*^{\mu\nu}(\widetilde k,\widehat l) 
    = \frac{e_f^2}{8} 
      \Tr \big[\widetilde\ksl \gamma^\nu \widehat\lsl \gamma^\mu \big] 
\end{align}
This is needed: (i) to ensure that $q_\mu H_*^{\mu\nu} = 0$, hence 
the gauge invariance of the hadronic tensor, and (ii) to allow use of
the Ward identities in proofs of factorization \cite{Collins:2007ph}. 
This approximation, made on the hard scattering coefficient,
is less critical then the kinematic approximations previously
discussed because it does not change in itself the kinematics of the
process. It is analogous to the approximation taken in considering the
usual handbag diagram of Fig.~\ref{fig:DIS_LO} 
with a massless quark line joining the 2 virtual photon, 
except that it approximates only the computation of the Dirac traces.

Finally, we define the LO hard scattering tensor
\begin{align}
  \HH_f^{\mu\nu}(\widetilde k,q,m_j^2)
    & = H_*^{\mu\nu}(\widetilde k,\widehat l) 
    \delta(\widetilde l^2-m_j^2) \\
  & = \Tr \big[\widetilde\ksl \gamma^\nu \widehat\lsl \gamma^\mu \big]
    \delta(\widetilde l^2-m_j^2) \ ,
\end{align}
which for $m_j^2=0$
coincides with the LO hard scattering tensor
computed for a diagram without jet function, as in
Eq.~\eqref{eq:pQCDfactnolims}. 
The hadronic tensor can then be written in factorized form as
\begin{align}
  W^{\mu\nu}(p,q) & =  \int_0^\infty\hspace*{-.3cm}dm_j^2\,J_2(m_j^2) 
 \label{eq:W-JET} \\
  & \times \int_\xi^{\xi/x_B} \frac{dx}{x}  
    \HH_f^{\mu\nu}(\widetilde k,q,m_j^2)\, \varphi_q(x,Q^2) \ ,
  \nonumber
\end{align}
which is the central result of this section. 
The transverse structure function reads
\begin{align}
\begin{split}
  & F_T(x_B,Q^2,m_N^2) =
    \int_0^\infty\hspace*{-.3cm}dm_j^2\,J_2(m_j^2) \\
  & \qquad\qquad \times  \sum_f \int_\xi^{\xi/x_B} \frac{dx}{x} 
  h_{f|T}(\widetilde x_f,Q^2,m_j^2) \varphi_q(x,Q^2) \ ,
  \label{eq:FT-JET} 
\end{split}
\end{align}
with $\varphi_q$ defined in Eq.~\eqref{eq:quarkPDFatLO}.
The longitudinal structure function $F_L=0$ because $h_L=0$.
An explicit computation gives $h_T(\widetilde x_f,Q^2) = \frac12 e_f^2
x \delta\big(x-\xi(1+m_j^2/Q^2)\big)$ so that at LO
\begin{align}
  & F_T(x_B,Q^2,m_N^2) 
 \label{eq:FTL_TMC_JMC_precise} \\
  & \qquad = \int_0^{\frac{1-x_B}{x_B}Q^2} \hspace*{-.2cm} dm_j^2 J_2(m_j^2) 
    F_T^{(0)}\bigg(\xi\Big(1+\frac{m_j^2}{Q^2}\Big),Q^2\bigg) \ , \nonumber 
\end{align}
Note that when $Q^2 \gg \vev{m_j^2}$, where $\vev{m_j^2} = \int dm_j^2\,
m_j^2 J_2(m_j^2)$, the massless $F_T^{(0)}$ decouples from the integration
over the jet mass, and we recover the TMC to the LO structure functions. 

\begin{figure*}[tb]
  \vspace*{0cm}
  \centerline{
  \includegraphics
    [width=0.45\linewidth]
    {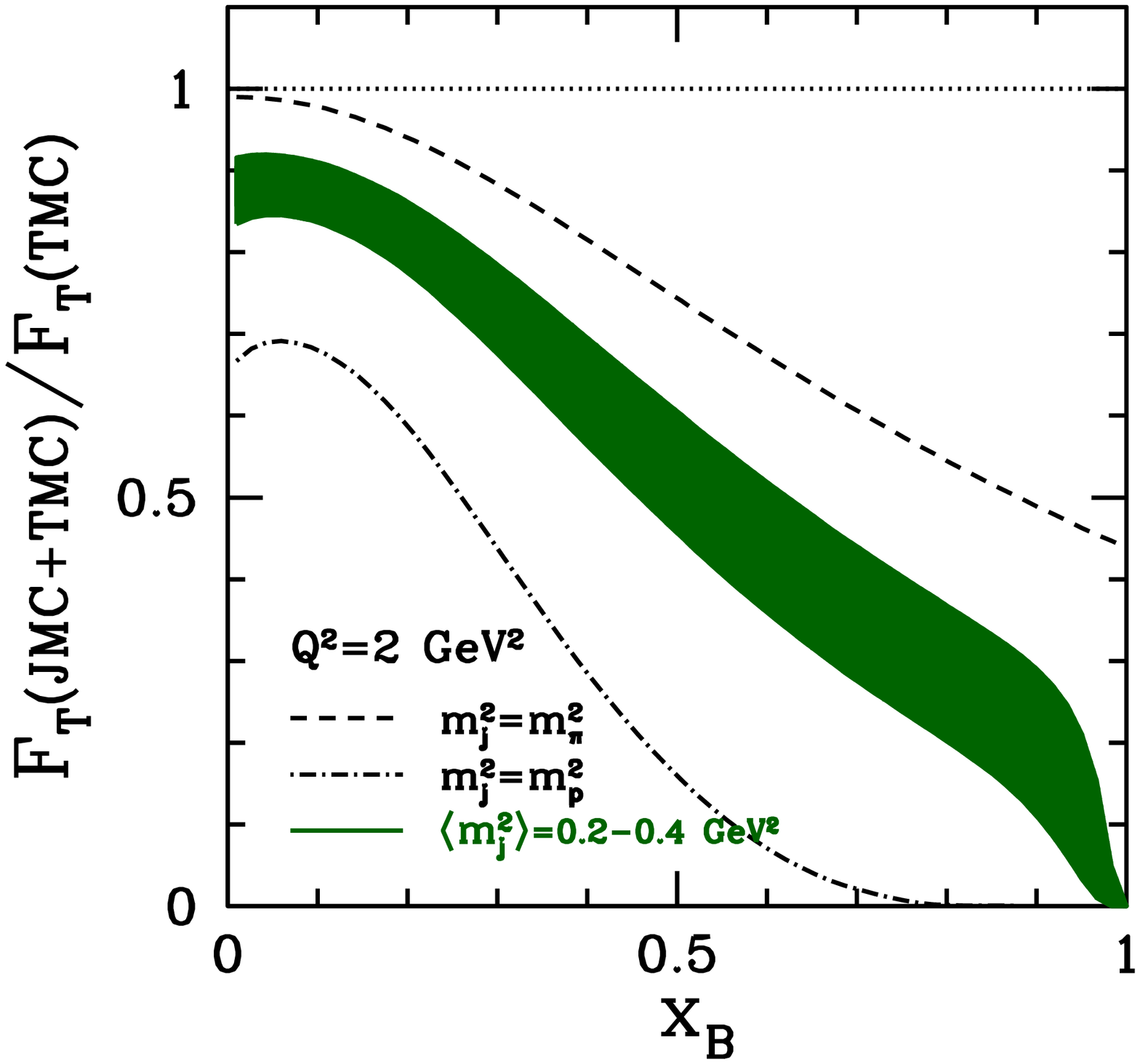}
  \hspace*{0.5cm}
  \includegraphics
    [width=0.45\linewidth]
    {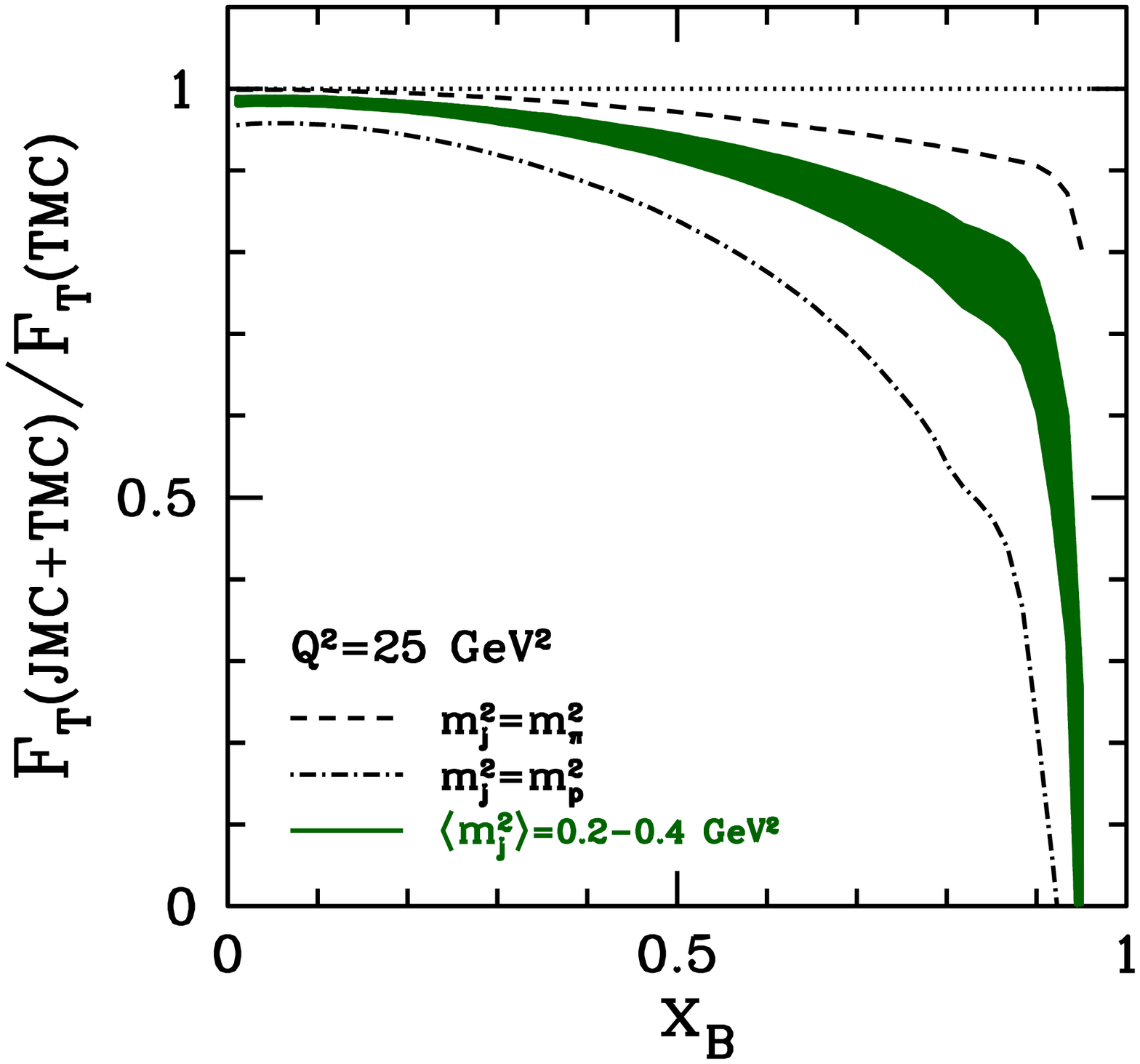}
  }
  \caption[]{Effect of jet mass corrections on $F_T$, computed with a
    toy jet spectral function as described in the
    text. Plotted is the ratio of $F_T$ with both TMC and JMC to $F_T$
    with only TMC included, as a function of $x_B$ for $Q^2=2$ and 25
    GeV$^2$. The shaded band corresponds to a log-normal jet mass
    distribution with $\vev{m_j^2} = 0.2-0.4$ GeV$^2$ and
    $\sigma_{m_j^2} = \vev{m_j^2}-2 \vev{m_j^2}$. The dashed and dot-dashed
    lines corresponds to delta functions at $m_j^2=m_\pi^2$ and
    $m_j^2=m_N^2$, respectively. 
  }
 \label{fig:FT_JMC}
\end{figure*}

\subsection{The jet spectral function}

Let us discuss more in detail the properties of the nonperturbative
quark spectral function $J_2$, defined in
Eq.~\eqref{eq:j2-spectral}. Let us start from the definition of the 
$j_2$ component of the jet function,  
\begin{align}
\begin{split}
   j_2(l) & = \frac{1}{4l^-}\Tr\big[ \gamma^- \hat J(l)\big] \\
     & = \sum_Y \delta^{(4)}\big(l-\sum_{i\in Y} p_i\big)
       \vev{0|\bar\psi_f(0)|Y}\gamma^- \vev{Y|\psi_f(0)|0} \ ,
\end{split}
\end{align}
with $f$ the quark flavor. For simplicity, we consider only light
quark flavors with $m_f \ll m_\pi$.
The color $c$ of the quark operator $\psi$ is not neutralized, so
that it must appear in the final state $\ket{Y}$. 
In the physical process, we are 
assuming that the struck quark's color is neutralized by a soft
gluon exchange with the target's remnant.
We also assume that we can neglect
the soft exchange for the purpose of evaluating the change of 
kinematics induced by the inclusion of a quark jet function on 
the lowest order contribution to the inclusive DIS. 
This assumption is likely valid if the jet and 
target rapidities are sufficiently separated.  
However, this might not be the case 
close to the kinematic limit $x_B=1$, and the approximation
will break down as we shall soon see. Because of color
confinement, we may assume that no more than 1 particle in the final
state is colored, all the other ones binding into colorless hadrons.
The colored particle must be a quark, to match the quark operator's
color, and we denote it by $\ket{q_{f'}^c}$.
Hence, the final state is made of 1 quark plus
an arbitrary number of hadrons, the lightest of which is a pion:
\begin{align}
  \ket{Y} = \ket{q_{f'}^c}\ket{h_1} \cdots \ket{h_N} \ ,
  \label{eq:finalstates}
\end{align}
with $N\geq0$ and $f'=f$ when $N=0$. 
The spectral function $J_2$, defined in Eq.~\eqref{eq:j2-spectral},
can be written as 
\begin{align}
  J_2(m_j^2) = Z \delta(m_j^2 - m_q^2) + (1-Z) \rho(m_j^2)  \ ,
  \label{eq:J2general}
\end{align}
where the $\delta$-function is due to the contribution of the single
particle $\ket{q_{f'}^c}$, and $0<Z<1$ \cite{Weinberg:1995mt}. The
continuous and positive definite function $\rho$ is the contribution
of  multiparticle states with $N>0$ in Eq.~\eqref{eq:finalstates} and
is normalized to 1 because of Eq.~\eqref{eq:J2norm}. Due to the
assumption \eqref{eq:finalstates}, $\rho$ has a bell shape:
it is equal to 0 up to $m_j^2 = (m_\pi+m_q)^2 \approx m_\pi^2$,
increases up to a maximum and then tends to 0 as $m_j^2 \ra \infty$ to
satisfy the normalization condition.
Using Eq.~\eqref{eq:J2general} in \eqref{eq:FTL_TMC_JMC_precise}, we
obtain 
\begin{align}
\begin{split}
  & F_T(x_B,Q^2,m_N^2) 
    = Z \, F_T^{(0)}\big(\xi,Q^2\big) \\
  & \quad + (1-Z) \int_{m_\pi}^{\frac{1-x_B}{x_B}Q^2} \hspace*{-.2cm} 
    dm_j^2 \rho(m_j^2) 
    F_T^{(0)}\bigg(\xi\Big(1+\frac{m_j^2}{Q^2}\Big),Q^2\bigg) \ .
  \label{eq:JMCnosoft}
\end{split}
\end{align}
Setting $Z=1$ is equivalent to calculating the standard handbag
diagram without the jet function, and one recovers the TMC formula.

The first term in Eq.~\eqref{eq:JMCnosoft} shows that the introduction
of the jet function in the handbag diagram goes some way toward
softening the problem with the unphysically positive $F_T$ at $x_B=1$,
but does not solve it. The reason is that we cannot kinematically
neglect the effect of the color neutralizing soft interactions when we
compute the hadbag diagram close to $x_B=1$, where the
rapidity difference between the current and target jets
is becoming smaller and smaller. A full solution
to this problem is the inclusion of a ``soft function'', in
addition to the target and jet functions, which describes the soft
exchanges in the context of fully unintegrated
correlation functions \cite{Collins:2007ph}. 
The soft function has essentially 
the effect of smearing the jet function, avoiding the
singular behavior displayed by the $\delta$-function. 
For a phenomenological inclusion of the soft function in collinear
factorization, we can substitute $J_2$ with a continuous function 
$J_m$ such that 
\begin{align}
  J_m(m_j^2) \xrightarrow[m_j^2 \ra 0]{} 0 \ ,
\end{align}
because of phase space, and
\begin{align}
  J_m(m_j^2) \xrightarrow[m_j^2 \gg m_\pi^2]{} J_2(m_j^2) \ .
\end{align}
It can be physically interpreted as the (smeared) jet mass distribution,
analogously to $J_2$, and we will call it smeared jet spectral
function. The structure function is then computed as in the Ansatz
discussed at the beginning of the Section: 
\begin{align}
  & F_T(x_B,Q^2,m_N^2) 
    \label{eq:JMCeffsoft} \\
  & \qquad = \int_{m_\pi}^{\frac{1-x_B}{x_B}Q^2} \hspace*{-.2cm} 
    dm_j^2 J_m(m_j^2) 
    F_T^{(0)}\bigg(\xi\Big(1+\frac{m_j^2}{Q^2}\Big),Q^2\bigg) \ .
    \nonumber
\end{align}
We note that the jet spectral function $J_2$ is defined as a quark
correlation function in vacuum, therefore it is process-independent.
On the other hand, the smeared jet function $J_m$ is process-dependent
because it effectively includes the soft momentum exchange with the
target. As a result, it's shape at $m_j^2 \lesssim m_\pi^2$ might
depend on $x_B$ and $Q^2$. However, the average jet mass squared,
$\vev{m_j^2}_m=\int_0^\infty dm_j^2 \, m_j^2\, J_m(m_j^2)$ should
exhibit a small sensitivity on $x_B$ and $Q^2$ because we may expect 
$\vev{m_j^2} \gg m_\pi^2$, see Section~\ref{sec:numest}.
Eq.~\eqref{eq:JMCeffsoft} is a reasonable approximation to the full
handbag diagram 
computation in the region of $(x_B,Q^2)$ phase space where the
integration over $dm_j^2$ in Eq.~\eqref{eq:FTL_TMC_JMC_precise}
extends well beyond the peak of the continuum $\rho(m_j^2)$, 
namely if 
\begin{align}
  \frac{1-x_B}{x_B}Q^2 \gtrsim \vev{m_j^2}_\rho \ ,
  \label{eq:JmJ2validity}
\end{align}
where
\begin{align}
  \vev{m_j^2}_\rho = \int_{m_\pi^2}^\infty dm_j^2 \, m_j^2
    J_2(m_j^2) \ .
\end{align}
In these conditions, the structure function \eqref{eq:JMCeffsoft} is
not much sensitive to the behavior of the jet function at small 
$m_j^2$, where $J_m$ may substantially differ from $J_2$.

For practical applications of JMC to global QCD fits of the PDFs, it
is necessary to develop a flexible enough and realistic
parametrization of the smeared jet spectral function $J_m$. For this
purpose, one may try to use a Monte Carlo simulation of DIS events in
order to generate enough data and test possible parametrizations. One
may also study the invariant jet mass distribution in $e^++e^- \ra
\,\text{jets}$ events, where the same jet function $\hat J$ discussed in
this Section appears in the LO cross-section. However, these studies
lie outside the scope of this paper, and we leave them for the
future. 

\subsection{Numerical estimates}
\label{sec:numest}
  
In order to obtain an estimate for the magnitude of JMC and of the
present theoretical uncertainty, we employ a toy model for the jet
spectral function. Let's consider a bell-shaped smooth function such
as the log-normal distribution 
\begin{align}
  f(x;\mu,\sigma) & = \frac{1}{x\sigma \sqrt{2\pi}}  
    \exp\left[ - \frac{(\log x - \mu)^2}{2\sigma^2} \right]
\end{align}
where 
\begin{align}
\begin{split}
  \mu & = \frac12 \log \left( \frac{\bar x^4}{\bar x^2+\sigma_x^2}
    \right) \\     
  \sigma & = \left[ \log\left( \frac{\sigma_x^2}{\bar x^2}+1
     \right)\right]^{\frac 12} \ ,
\end{split}
\end{align}
and $\bar x$ and $\sigma_x$ are the average value of $x$ and its
standard deviation. Then, we can parametrize the continuum
part $\rho$ of the toy jet mass distribution in
Eq.~\eqref{eq:FTL_TMC_JMC_precise}  
in terms of the average jet mass $\vev{m_j^2}_\rho$ and its standard
deviation $\sigma_{m_j^2}$:
\begin{align}
  \rho(m_j^2) = f(m_j^2-m_\pi^2;\mu,\sigma) \ ,
\end{align}
with 
\begin{align}
\begin{split}
  \mu & = \vev{m_j^2}_\rho - m_\pi^2 \\    
  \sigma & = \sigma_{m_j^2} 
\end{split}
\end{align}
in units of GeV$^2$. 
From the typical particle 
multiplicity of the current jet at the JLab energy, 
we estimate $\vev{m_j^2}_\rho =0.2-0.4$ GeV$^2$, and assume 
$\sigma_{m_j^2} = C \vev{m_j^2}_\rho$ with $C=1-2$.

In Fig.~\ref{fig:FTvsFT0}, we plot the JMC to the transverse structure
function as obtained in Eq.~\eqref{eq:FTL_TMC_JMC_precise} by
neglecting soft momentum exchanges.
The dashed line corresponds to $Z=1$, and is equivalent to computing
only TMC. The solid line corresponds to JMC coming only from the
continuum part $\rho$ of the jet spectral function, i.e., $Z=0$.
For comparison, the massless structure function is plotted as a dotted
line. The true jet mass corrected $F_T$ should lie somewhere in
between because $0<Z<1$, in general. With the smearing due to soft
interactions, see Eq.~\eqref{eq:JMCeffsoft}, the true 
$F_T$ will tend to 0 as $x_B\ra1$.

The sensitivity of JMC to the $\vev{m_j^2}_\rho$ value can be gauged from 
Fig.~\ref{fig:FT_JMC}, where we plotted the ratio of the $Z=1$ TMC-only
structure function to the $Z=0$ jet mass corrected structure function.
For comparison, we also use $\rho(m_j^2)=\delta(m_j^2-m_\pi^2)$ and
$\rho(m_j^2)=\delta(m_j^2-m_N^2)$, considered as extreme cases of JMC. 
In the absence of a better knowledge of the value of $Z$ and
$\vev{m_j^2}$, the overall theoretical uncertainty on JMC can be quite
large, especially at low $Q^2=2$ GeV$^2$, and is still
non-negligible at $Q^2=25$ GeV$^2$. At moderate $x_B \lesssim 0.6$
it is of the same order of magnitude of the TMC corrections to the
massless $F_T$. 

\begin{figure}[t]
  \vspace*{0cm}
  \centerline{
  \includegraphics
    [width=0.9\linewidth]
    {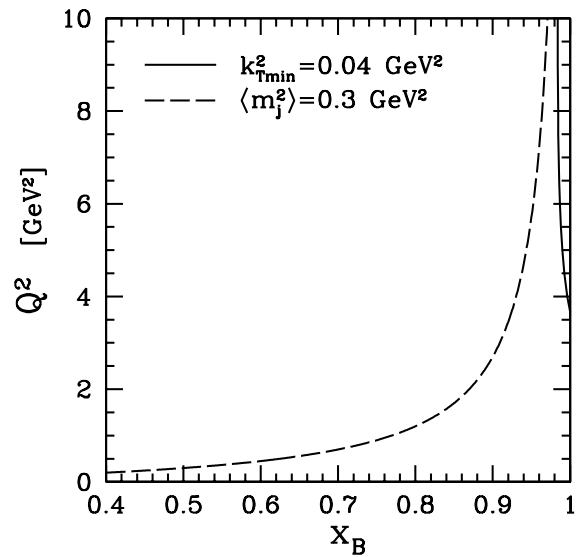}
  }
  \vskip-.5cm
  \caption[]{
    Range of validity of kinematic approximations used in deriving TMC
    and JMC. The solid line corresponds to Eq.~\eqref{eq:mjlimits} with
    $\vev{m_j^2}=0.3$ GeV$^2$; the dashed line corresponds to
    Eq.~\eqref{eq:ktlimits} and \eqref{eq:kTaverage}, with
    $k_{T\text{min}}^2=0.04$ GeV$^2$. 
  }
 \label{fig:confidenceregion}
\end{figure}

Finally, we want to estimate in which region of $x_B$ and $Q^2$
the kinematic approximations involved in step 3 and 4 of the
factorization procedure are expected to be valid.
As we discussed after Eq.~\eqref{eq:W-step3}, replacing $j_2(l)$ with 
$j_2(\widetilde l)$ makes sense only if the integral over $dm_j^2$ 
is dominated by $m_j^2 \approx m_{j|\text{max}}^2$, where the spectral
function $J_2$ has a maximum, hence 
minimal slope. Looking at the integration limits in
Eq.~\eqref{eq:JMCeffsoft}, and noticing that 
for a probability distribution with the properties of the jet
spectral function it is typically true that 
$m_{j|\text{max}}^2 \lesssim \vev{m_j^2} \lesssim \vev{m_j^2}_\rho$, where
$\vev{m_j^2} = \int_0^\infty dm_j^2\, m_j^2\, J_m(m_j^2)$,  we obtain
the following condition:
\begin{align}
  \frac{1-x_B}{x_B} Q^2 \gtrsim \vev{m_j^2}_\rho \ .
 \label{eq:mjlimits}
\end{align}
Note that it coincides with the condition \eqref{eq:JmJ2validity} that
insures we can indeed approximate $J_2 \approx J_m$ for the
computation of inclusive DIS cross section. It also guarantees some
rapidity separation between the current and target jets, which is
needed to justify the handbag diagram in the first place.
The approximation of step 4 consisted in neglecting the integration
limits on $dk^-$ and $d^2k_T$. For the transverse momentum, 
we need at least to make sure that the average $\vev{k_T^2}$ is well
below the upper limit derived in Eq.~\eqref{eq:ktbound}:
\begin{align}
  \vev{k_T^2} \ll \frac{1-\xi}{4\xi} Q^2 
    \big( 1 + \xi\frac{m_N^2}{Q^2} \big) \ .
 \label{eq:ktlimits}
\end{align}
The difficulty is that we cannot estimate $\vev{k_T^2}$ within 
collinear factorization. To do this, we would need to resort to unintegrated
PDF \cite{Collins:1981uw,Kimber:2001sc}, which are still 
integrated over $dk^-$, or to the more recently proposed fully
unintegrated PDF \cite{Collins:2007ph}.  
For a rough estimate, we may use the uncertainty principle and set the
minimum transverse momentum $k_{T\text{min}}^2 = 1/r_N^2 \approx
0.04$ GeV$^2$, where $r_N^2$ is the nucleon radius. pQCD evolution
will then broaden it roughly according 
to 
\begin{align}
  \vev{k_T^2} = k_{T\text{min}}^2  [1+ C \log(Q^2/k_{T\text{min}}^2)]
  \ ,  
 \label{eq:kTaverage}
\end{align}
with $C$ a constant of order 1. 
The borders of the confidence region for the discussed TMC and JMC are
plotted in Fig.~\ref{fig:confidenceregion} using the above estimate
for $\vev{k_T^2}$, and $\vev{m_j^2}=0.3$ GeV$^2$.


\section{Summary and conclusions}
\label{sec:conclusions}

In the first part of this paper, 
we computed the target mass corrections to unpolarized 
DIS structure functions in the context of collinear factorization.
Because of the non-perturbative nature of target mass, 
we emphasized that for any factorizable hadronic observable, 
the TMC can only appear explicitly in kinematic variables and 
implicitly in definitions of non-perturbative hadron matrix elements.  
The momentum space approach allowed us to avoid the ambiguities
related to the moments inversion which affect the OPE treatment of
Georgi and Politzer. In particular, we could respect 4-momentum and
baryon number conservation, and obtain TMC corrected structure
functions without unphysical contributions at $x_B>1$. 
When performing global QCD fits of the PDFs
in the context of pQCD collinear factorization, the procedure
presented in this paper might be the most consistent way to treat TMC, 
because it expresses the long distance physics of structure functions,
and the leading target mass correction, in terms of PDFs that share the
same partonic operators with the PDFs of zero hadron mass. Hence it
allows to unambiguously separate the kinematic effects of the target's
mass from its dynamical contribution to parton matrix elements and the
PDFs. 

Our formalism for TMC in Eq.~\eqref{eq:FTL_TMC}
is valid at leading twist and any order in
$\alpha_s$. Calculating TMC for the power-suppressed higher-twist
contributions to the structure functions is a non-trivial
\cite{Ellis:1982cd} but important issue for measuring the size 
of parton correlations in the nucleon wave-function, which we leave
to a future effort. The leading-twist formalism can be easily
extended to polarized DIS structure functions \cite{AW}, for which a
correct evaluation of TMC is even more important than in the
unpolarized case because the bulk of available data is in fact in the
large-$x_B$ domain.  
The extension to semi-inclusive DIS and to hadronic collisions is also
very important, in order to fully include TMC in global PDF fits. An example is
the Drell-Yan cross-section at large Feynman $x_F$, which has the
potential to further constrain large-$x$ PDFs \cite{Owens:2007kp}. 
It is also straightforward to extend the TMC analysis 
to DIS on nuclear targets, in order to include the effects
of nucleon binding and Fermi motion \cite{AQV}. This is especially
important for studying the large-$x$ neutron PDFs and the $d/u$ ratio,
which are extracted from data taken with a Deuterium target.

In the second part of the paper, we examined the impact of a
final-state jet function on the extraction of PDFs at large $x_B$. 
We proposed to write the leading order hadronic tensor, 
hence the lowest order contribution to DIS cross section,
in terms of the spectral representation $J_2$ of the jet function,
which has the physical meaning of invariant jet mass distribution. 
We evaluated the impact of JMC on the leading order DIS
structure functions, and found it to be 
potentially large even at not so small values of photon virtuality
such as $Q^2 = 25$ GeV$^2$. 
In the NLO cross-section, the impact of JMC is likely to be
reduced, because a non-zero jet invariant mass can be produced in the
hard scattering beyond tree level, but is still potentially large.
We also evaluated the range of validity in
$x_B$ and $Q^2$ of the approximations we made.

For practical applications to global fits of PDFs, it is important
to investigate the shape and properties of the smeared jet spectral
function $J_m$, which effectively includes the neglected soft momentum
exchanges in the final state.
This can be phenomenologically done using a Monte-Carlo simulation and
then trying several parametrizations of $J_m$. In a more fundamental
approach, we noticed that the jet spectral function $J_2$ is related 
to the non-perturbative quark propagator, which can be computed in
lattice QCD or using Schwinger-Dyson equations. To avoid the
difficulties connected to the analytic continuation to
Minkowski space, one may try and rotate the whole handbag
diagram to Euclidean space, or use a Hamiltonian-based formulation of
lattice QCD.

In conclusion, the obtained results on TMC and JMC will be very important
when using large-$x_B$ and low-$Q^2$ data on DIS structure
function (like those obtained at Jefferson Lab)
to extract reliable PDFs at large-$x$, and to disentangle
kinematic effects from the dynamically interesting higher-twist parton
correlations. The discussed extensions of our formalism to other
processes will allow a full inclusion of TMC and JMC in global QCD
fits of parton distribution functions.


\begin{acknowledgments}
We thank C.~Aubin, T.~Blum, M.~E.~Christy, J.~C.~Collins, V.~Guzey,
C.~E.~Keppel, P.~Maris, W.~Melnitchouk, D.~Richards, C.~Weiss
for useful discussions and suggestions. This work has been supported
in part by the U.S. Department of Energy, Office of 
Nuclear Physics, under contract DE-AC02-06CH11357
and contract DE-AC05-06OR23177 
under which Jefferson Science Associates, LLC     
operates the Thomas Jefferson National Accelerator Facility, and 
under grant DE-FG02-87ER40371 and 
contract DE-AC02-98CH10886.
\end{acknowledgments}

\begin{appendix}

\section{Kinematic constraints at finite $Q^2$}
\label{app:kinematics}

Let us consider the handbag diagram for a DIS process on a nucleon
target, as depicted in the right hand side of 
Fig.~\ref{fig:DISfactorization}. We repeat the kinematic analysis of
the handbag diagram performed in Section~\ref{sec:TMC}, but for the
general case of an off-shell bound parton of momentum $k$, and $k^2 \lesssim
m_f^2$. The limit of on-shell quarks of mass $m_f^2$,
relevant to collinear factorization, can be obtained setting
$k^2=m_f^2$ and $x_f = \widetilde x_f$ in the formulae below. 

\begin{figure}[tb]
  \vspace*{0cm}
  \centerline{
  \includegraphics
    [width=0.9\linewidth]
    {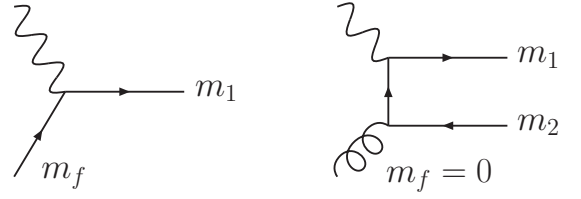}
  }
  \caption[]{
    Lowest order couplings of a generic vector boson ($\gamma, W^\pm,
    Z$) to a parton of flavor $f$ and mass $m_f$. The masses of the
    produced quarks are $m_1$ and $m_2$. Left: boson-quark
    scattering ($m_2=0$). Right: boson-gluon fusion. 
  }
 \label{fig:LOcouplings}
\end{figure}

We consider the scattering of a
generic vector boson ($\gamma, W^\pm, Z$) on a parton of flavor $f$
of mass $m_f$. The lowest order couplings are displayed in
Fig.~\ref{fig:LOcouplings}. The masses of the quarks (other than $f$)
coupled to the vector boson are $m_1$ and $m_2$. The current jet mass
must satisfy  
\begin{align}
  m_j^2 \geq \sth
\end{align}
where
\begin{align}
  \sth = (m_1+m_2)^2 \ .
\end{align}
As discussed in Sec.~\ref{sec:TMC}, the net baryon number is likely 
to flow through the bottom of the handbag diagram for the leading 
DIS contribution that is given by the collinear factorization formalism.
Therefore,
\begin{align}
  \sth \leq m_j^2 \leq s-m_N^2 \ .
\end{align}
Using $m_j^2 = (q+k)^2 = k^2 + (1/x_f-1)Q^2$ we obtain
\begin{align}
  \frac{x_B}{1-x_B k^2/Q^2} 
  \leq x_f \leq 
  \frac{1}{1+(\sth-k^2)/Q^2} \ .
 \label{eq:xflims}
\end{align}
Using $m_j^2=(Q^2+\frac{\xi}{x}k^2)(\frac{x}{\xi}-1)$,
Eq.~\eqref{eq:xflims} can alternatively be expressed as limits over the
fractional momentum $x=k^+/p^+$: 
\begin{align}
  x^{min} \leq x \leq x^{max}
\end{align}
where
\begin{align}
  & x^{min} = \xi \frac{Q^2+\sth-k^2+\Delta[k^2,-Q^2,\sth]}{Q^2} 
    \nonumber \\
  & x^{max} = \xi \frac{Q^2+s-m_N^2-k^2+\Delta[k^2,-Q^2,s-m_N^2]}{Q^2} 
     \nonumber \\
  & \Delta[a,b,c] = \sqrt{a^2 + b^2 + c^2 - 2(ab+bc+ca)} \ .
 \label{eq:xlims}
\end{align}
We finally note that
\begin{align}
  x_f & = \frac{\xi}{x} 
  \frac{1}{1-\frac{\xi^2}{x^2}\frac{k^2}{Q^2}} \ .
\end{align}

\section{Invariant and helicity structure functions}
\label{app:strfns}

\subsection{Helicity structure functions}
\label{app:helicitystrfns}

We work in a collinear frame and for generality we keep the quark mass
different from zero.
For collinear on-shell partons we have from Eq.~\eqref{eq:kinematics}
\begin{align}
\begin{split}
  p^\mu & = p^+ \nbar^\mu 
          + \frac{m_N^2}{2 p_A^+} n^\mu \\
  q^\mu & = - \xi p^+ \nbar^\mu 
          + \frac{Q^2}{2\xi p^+} n^\mu \\
  \widetilde k^\mu & = x p^+ \nbar^\mu 
          + \frac{m_f^2}{2xp^+} n^\mu \ ,
\end{split}
\end{align}
where $m_f$ is the mass of the parton of flavor $f$. For later use, we define the shorthands
\begin{align}
  \rho_B^2 = 1+4x_B^2\frac{m_N^2}{Q^2} \qquad 
  \rho_f^2 = 1+4x_f^2\frac{m_f^2}{Q^2} \ ,
\end{align}
where, as in Eq.~\eqref{eq:invariants},
\begin{align}
    x_B = \frac{-q^2}{2 p\cdot q} \qquad
    x_f = \frac{-q^2}{2 \widetilde k\cdot q } \ .
\end{align}

Following \cite{Aivazis:1993kh}, we define the longitudinal, transverse and
scalar polarization vectors with respect to the virtual photon
momentum $q$ and a reference vector $p$,
\begin{align}
  \varepsilon_0^\mu(p,q) 
    & = \frac{-q^2p^\mu+(p\cdot q)q^\mu}{\sqrt{-q^2}[(p\cdot q)^2-q^2p^2]}
    = \frac{-q^2p^\mu+(p\cdot q)q^\mu}{\sqrt{-q^2}(p\cdot q) \rho^2(p,q)} 
    \nonumber \\
  \varepsilon_\pm^\mu(p,q) 
    & = \frac{1}{\sqrt{2}} (0,\pm 1,-i,0) \nonumber \\
  \varepsilon_q^\mu(p,q) 
    & = \frac{q^\mu}{\sqrt{-q^2}} \ ,
 \label{eq:polvect}
\end{align}
where
\begin{align}
  \rho^2(p,q) = 1-p^2q^2/(p\cdot q)^2 \ .
\end{align}
It is immediate to verify that $\rho^2(p,q) = \rho_B^2$ and
$\rho^2(\widetilde k,q) = \rho_f^2$.
The polarization vectors satisfy the following conditions
\begin{alignat}{2}
  & \varepsilon_\lambda \cdot \varepsilon_{\lambda'} = 0 
    & \qquad & \text{for\ } \lambda\neq\lambda' \nonumber \\
  & \varepsilon_\lambda \cdot \varepsilon_\lambda = 1 
    & \qquad & \text{for\ } \lambda = 0,+,- \\
  &\varepsilon_q \cdot \varepsilon_q = - 1 \nonumber
\end{alignat}
and, in particular, $q\cdot \varepsilon_0 = q\cdot \varepsilon_\pm = 0$.
The helicity structure functions $F_\lambda$ are defined as
projections of the hadronic tensor:
\begin{align}
  F_\lambda(x_B,Q^2) = P_\lambda^{\mu\nu}(p,q) W_{\mu\nu}(p,q)
\end{align}
with $\lambda=L,T,A,S,\{0q\},[0q]$. The longitudinal, transverse,
axial, scalar, and mixed projectors $P_\lambda^{\mu\nu}$ are
\begin{align}
\begin{split}
  P_L^{\mu\nu}(p,q) & = \varepsilon_0^\mu(p,q) \varepsilon_0^{\nu*}(p,q) \\
  P_T^{\mu\nu}(p,q) & = \varepsilon_+^\mu(p,q) \varepsilon_+^{\nu*}(p,q) 
    + \varepsilon_-^\mu(p,q) \varepsilon_-^{\nu*}(p,q) \\
  P_A^{\mu\nu}(p,q) & = \varepsilon_+^\mu(p,q) \varepsilon_+^{\nu*}(p,q) 
    - \varepsilon_-^\mu(p,q) \varepsilon_-^{\nu*}(p,q) \\
  P_q^{\mu\nu}(p,q) & = \varepsilon_q^\mu(p,q) \varepsilon_q^{\nu*}(p,q) \\
  P_{\!\!\{0q\!\}}^{\mu\nu}(p,q) 
    & = \varepsilon_0^\mu(p,q) \varepsilon_q^{\nu*}(p,q)
    + \varepsilon_q^\mu(p,q) \varepsilon_0^{\nu*}(p,q) \\
  P_{[0q]}^{\mu\nu}(p,q) & = \varepsilon_0^\mu(p,q) \varepsilon_q^{\nu*}(p,q)
    - \varepsilon_q^\mu(p,q) \varepsilon_0^{\nu*}(p,q) \ . 
 \label{eq:helicityprojectors}
\end{split}
\end{align}
Using 
\begin{align}
\begin{split}
  & \varepsilon_+^\mu(p,q) \varepsilon_+^{\nu*}(p,q) 
    - \varepsilon_-^\mu(p,q) \varepsilon_-^{\nu*}(p,q)
    = \frac{-i\varepsilon^{\mu\nu\alpha\beta} p_\alpha q_\beta}
      {(p\cdot q) \rho_B} \\
  & \varepsilon_+^\mu(p,q) \varepsilon_+^{\nu*}(p,q) 
    + \varepsilon_-^\mu(p,q) \varepsilon_-^{\nu*}(p,q) \\
  & \qquad= -g^{\mu\nu} + \varepsilon_0^\mu(p,q) \varepsilon_0^{\nu*}(p,q)
    - \varepsilon_q^\mu(p,q) \varepsilon_q^{\nu*}(p,q) \ , 
\end{split}
\end{align}
one easily sees that 
\begin{align}
  F_T(x_B,Q^2) & = - W^\mu_\mu(p,q) + F_L(x_B,Q^2) - F_q(x_B,Q^2) 
    \nonumber \\
  F_A(x_B,Q^2) & = \frac{-i\varepsilon^{\mu\nu\alpha\beta} p_\alpha q_\beta}
      {(p\cdot q) \rho_B} W_{\mu\nu}(p,q) \ .
 \label{eq:helicityids}
\end{align}
Even if not apparent from Eq.~\eqref{eq:polvect}, a consequence of the
normalization conditions is that the reference vector has the only
function to define the $t-z$ and transverse planes in conjunction
with $q^\mu$: as long as it lays in the $t-z$ plane, a different
reference vector defines the same polarization vectors
\cite{Aivazis:1993kh}. For example, $\varepsilon_\lambda^\mu(p,q) =
\varepsilon_\lambda^\mu(\widetilde k,q)$.
As we will see, choosing $\widetilde k$ instead of $p$ is convenient when
defining the parton level helicity structure functions, which read
\begin{align}
  h_\lambda(x_f,Q^2) = P_\lambda^{\mu\nu}(\widetilde k,q) 
    \HH_{\mu\nu}(\widetilde k,q)
\end{align}
and satisfy identities analogous to Eq.~\eqref{eq:helicityids}, with $p
\ra \widetilde k$.

\subsection{Invariant structure functions}
\label{app:invsfn}

For a generic lepton-hadron scattering, 
we define the hadronic $F_i$ and partonic $h_i$ 
invariant structure functions with $i=1,\ldots,6$ by the following
tensor decomposition of the hadronic tensor:
\begin{align}
\begin{split}
  & W^{\mu\nu}(p,q) 
    = \Big( -g^{\mu\nu} + \frac{q^\mu q^\nu}{q^2} \Big) F_1(x_B,Q^2) \\
  & \quad + \Big(p^\mu - q^\mu\frac{p\cdot q}{q^2}\Big)
      \Big(p^\nu - q^\nu\frac{p\cdot q}{q^2}\Big)
      \frac{F_2(x_B,Q^2)}{p\cdot q} \\
  & \quad - i \varepsilon^{\mu\nu\alpha\beta} p_\alpha q_\beta 
      \frac{F_3(x_B,Q^2)}{p\cdot q} 
    - \frac{q^\mu q^\nu}{q^2} F_4(x_B,Q^2) \\
  & \quad  
    - \frac{p^\mu q^\nu + q^\mu p^\nu}{2 p\cdot q} F_5(x_B,Q^2) 
    + \frac{p^\mu q^\nu - q^\mu p^\nu}{2 p\cdot q} F_6(x_B,Q^2)
 \label{eq:invariantF}
\end{split}
\end{align}
and
\begin{align}
\begin{split}
  & \HH^{\mu\nu}(\widetilde k,q) 
    = \Big( -g^{\mu\nu} + \frac{q^\mu q^\nu}{q^2} \Big) h_1(\widetilde x_f,Q^2) \\
  & + \Big(\widetilde k^\mu - q^\mu\frac{\widetilde k\cdot q}{q^2}\Big)
      \Big(\widetilde k^\nu - q^\nu\frac{\widetilde k\cdot q}{q^2}\Big)
      \frac{h_2(\widetilde x_f,Q^2)}{\widetilde k\cdot q} \\
  & - i \varepsilon^{\mu\nu\alpha\beta} \widetilde k_\alpha q_\beta 
      \frac{h_3(\widetilde x_f,Q^2)}{\widetilde k\cdot q} 
    - \frac{q^\mu q^\nu}{q^2} h_4(\widetilde x_f,Q^2) \\
  & - \frac{\widetilde k^\mu q^\nu + q^\mu \widetilde k^\nu}
    {2 \widetilde k\cdot q} h_5(\widetilde x_f,Q^2) 
    + \frac{\widetilde k^\mu q^\nu - q^\mu \widetilde k^\nu}
    {2 \widetilde k\cdot q} h_6(\widetilde x_f,Q^2) \ .
 \label{eq:invarianth}
\end{split}
\end{align}
These 2 definitions differ from the notation of 
Ref.~\cite{Aivazis:1993kh} in
the chosen denominators. Our definitions have the advantage of
displaying a duality between the hadron and parton level, which
can be obtained from each other by exchanging $p \leftrightarrow
\widetilde k$, and lead to
a lesser degree of mixing between the hadron and parton structure
functions under collinear factorization, see Eq.~\eqref{eq:Fi_hi}.
By applying the projectors \eqref{eq:helicityprojectors} 
to Eqs.~\eqref{eq:invariantF}-\eqref{eq:invarianth}, 
it is straightforward to show that
\begin{xalignat}{2}
  &F_L = -F_1 + \frac{\rho_B^2}{2x_B} F_2 & \quad
    &h_L = -h_1 + \frac{\rho_f^2}{2\widetilde x_f} h_2 \nonumber \\
  &F_T = 2F_1 &
    &h_T = 2h_1 \nonumber\\
  &F_A = \rho_B F_3 &
    &h_A = \rho_f h_ 3 \nonumber\\
  &F_S = F_4 - F_5 &
    &h_S = h_4 - h_5 \nonumber\\
  &F_{\{\!0q\}}\! = -\rho_B F_5 &
    &h_{\{\!0q\}}\! = -\rho_f h_5 \nonumber \\ 
  &F_{[0q]} = -\rho_B F_6 &
    &h_{[0q]} = -\rho_f h_6 \ ,
 \label{eq:helvsinv}
\end{xalignat}
where we understood the dependence of $F_{i\lambda}$ on $(x_B,Q^2)$
and of $h_{i\lambda}$ on $(\widetilde x_f,Q^2)$ for ease of notation.
Note that $F_L$ differs by a factor of $2x_B$
with respect to other common conventions. In our notation, the ratio $R$
of transverse and longitudinal electron-nucleon cross sections reads
\begin{align}
  R = \frac{\sigma_T}{\sigma_L} = \frac{F_L}{F_1} \ .
\end{align}

\subsection{Collinear factorization for structure functions}
\label{app:colfactsfn}

As discussed in Section~\ref{sec:TMC} and
Appendix~\ref{app:kinematics}, the collinear factorization theorem
states that  
\begin{align}
\begin{split}
  W^{\mu\nu}(p,q) & = \sum_f \int \frac{dx}{x} \, 
    \theta(\widetilde x_f^{max}-\widetilde x_f) 
    \theta(\widetilde x_f-\widetilde x_f^{min})  \\ 
  & \times \HH_f^{\mu\nu}(\widetilde k,q) \,  \varphi_{f/N}(x,Q^2) 
 \label{eq:pQCDfactlims}
\end{split}
\end{align}
where
\begin{align}
  \widetilde x_f & = \frac{\xi}{x} 
    \frac{1}{1-\frac{\xi^2}{x^2}\frac{m_f^2}{Q^2}} \\
  \widetilde x_f^\text{min} & = \frac{x_B}{1-x_Bm_f^2/Q^2} \\  
  \widetilde x_f^\text{max} & = \frac{1}{1+(\sth-m_f^2)/Q^2} \ .
\end{align}
The corresponding limits of integration on $dx$, namely
$x^\text{min}$ and $x^\text{max}$, can be read off
Eq.~\eqref{eq:xlims} setting $k^2=m_f^2$.
As discussed in Appendix~\ref{app:helicitystrfns},
$P_\lambda^{\mu\nu}(p,q) = P_\lambda^{\mu\nu}(\widetilde k,q)$, hence
the factorization theorem for helicity structure functions reads
\begin{align}
\begin{split}
  & F_\lambda(x_B,Q^2,m_N^2) \\
  & \quad= \sum_f \int_{x^\text{min}}^{x^\text{max}} 
    \frac{dx}{x} \, h^f_\lambda(\widetilde x_f,Q^2) \,
    \varphi_{f/N}(x,Q^2) \\
  & \quad = \sum_f \int_{\widetilde x_f^\text{min}}^{\widetilde x_f^\text{max}} 
    \frac{d\widetilde x_f}{\widetilde x_f} \, h^f_\lambda(\widetilde x_f,Q^2) \,
    \varphi_{f/N}\Big(\frac{\xi}{\xi_f},Q^2\Big) \ ,
  \label{eq:helF_TMC}
\end{split}
\end{align}
where
\begin{align}
  \xi_f = \frac{2\widetilde x_f}
    {1+\sqrt{1+4\widetilde x_f^2m_f^2/Q^2}} \ .
\end{align}
The last line of Eq.~\eqref{eq:helF_TMC} is particularly interesting,
because the Nachtmann variable $\xi$ only appears in the argument of
$\varphi$, without touching the integration limits.
In shorthand notation, where we highlight the dependence on $x_B$ and
$\xi$ and suppress that on $m_N^2$ and $Q^2$, and understand the sum
over $f$, the helicity structure functions read 
\begin{align}
  F_\lambda(x_B) & \equiv h_\lambda^f \otimes \varphi_{f/N} (\xi) \ .
  \label{eq:Flambda_hi}
\end{align}
For the invariant structure functions, kinematic prefactors often
appear: 
\begin{align}
\begin{split}
  F_1(x_B) & = h_1^f \otimes \varphi_{f/N}(\xi) \\  
  F_2(x_B) & = \frac{x_B}{\widetilde x_f} \frac{\rho_f^2}{\rho_B^2}
    h_2^f \otimes \varphi_{f/N}(\xi)\\  
  F_{3,5,6}(x_B) & = \frac{\rho_f}{\rho_B} h_{3,5,6}^f 
    \otimes \varphi_{f/N}(\xi) \\
  F_4(x_B) & = h_4^f \otimes \varphi_{f/N}(\xi) 
    + \big( \frac{\rho_f}{\rho_B} - 1 \big) h_5^f \otimes \varphi_{f/N}(\xi) \ .
 \label{eq:Fi_hi}
\end{split}
\end{align}

The ``massless structure functions'' can be obtained by setting $m_N^2=0$,
hence, $\xi=x_B$ in Eqs.~\eqref{eq:Flambda_hi}-\eqref{eq:Fi_hi}:
\begin{align}
  F_{\lambda,i}^{(0)}(x_B) = F_{\lambda,i}(x_B)|_{m_N^2=0} \ .
\end{align}
In this definition we left the quark mass $m_f$ arbitrary.

The ``na\"ive'' target mass corrected structure functions
$F^\text{nv}$ are obtained by using $x\leq 1$ as an upper limit of 
integration over $dx$ in Eq.~\eqref{eq:helF_TMC}.  
This limit is a general and process-independent consequence of the
definition of a parton distribution in the field theoretic parton
model \cite{Jaffe:1983hp}, but in DIS it is weaker than $x \leq
x^\text{max}$, which is induced by 4-momentum and baryon number
conservation as discussed in Section~\ref{sec:TMC}. In detail, the
na\"ive helicity structure functions read   
\begin{align}
  & F_\lambda^\text{nv}(x_B) = \\
    & \quad= \sum_f \int_{x^\text{min}}^1 
    \frac{dx}{x} \, h_\lambda(\widetilde x_f,Q^2) \,
    \varphi_{f/N}(x,Q^2) 
  \label{eq:Fnv_hi}
\end{align}
Using the definition of massless structure functions, one finds
\begin{align}
\begin{split}
  F_{1,\lambda}^\text{nv}(x_B) & = F_{1,\lambda}^{(0)}(\xi) \\
  F_2^\text{nv}(x_B) & = \frac{1}{\rho_B^2} \frac{x_B}{\xi}
    F_2^{(0)}(\xi) \\
  F_{3,4,5}^\text{nv}(x_B) & = \frac{1}{\rho_B} F_{3,4,5}^{(0)}(\xi) \\
  F_4^\text{nv}(x_B) & = F_4^{(0)}(\xi) 
    + \frac{1-\rho_B}{\rho_B} F_5^{(0)}(\xi) \ . \\
\end{split}
\end{align}
These formulae have already appeared in
\cite{Aivazis:1993kh,Kretzer:2003iu,Kretzer:2002fr}, modulo the change
of notation discussed in Appendix~\ref{app:invsfn}. As already noted
in the main text, they are non-zero in the unphysical region $x_B>1$.

\subsection{Structure functions with Jet Mass Corrections}

At LO, the helicity structure functions with jet mass corrections read
\begin{align}
\begin{split}
  & F_\lambda(x_B,Q^2,m_N^2) \\
  & \qquad = \int_0^{\frac{1-x_B}{x_B}Q^2} \hspace*{-.2cm} dm_j^2 J_2(m_j^2) 
    F_\lambda^{(0)}\big(\xi(1+m_j^2/Q^2),Q^2\big) \ .
  \label{eq:FhelJMC}
\end{split}
\end{align}
The JMC to invariant structure functions can be obtained from
Eqs.~\eqref{eq:FhelJMC} and \eqref{eq:helvsinv}. Suppressing the $Q^2$
and $m_N^2$ dependence of the structure functions for ease of
notation, we obtain: 
\begin{align}
\begin{split}
  F_1^\text{JMC}(x_B) 
    & = \int_0^{\frac{1-x_B}{x_B}Q^2} \hspace*{-.2cm} dm_j^2 J_2(m_j^2) 
    F_1^{(0)}\big(\xi(1+m_j^2/Q^2)\big) \\
  F_2^\text{JMC}(x_B) 
    & = \int_0^{\frac{1-x_B}{x_B}Q^2} \hspace*{-.2cm} dm_j^2 J_2(m_j^2) 
    \frac{1}{\rho_B^2} \frac{x_B}{\xi(1+m_j^2/Q^2)} \\
    & \times  F_2^{(0)}\big(\xi(1+m_j^2/Q^2)) \\
  F_{3,5,6}^\text{JMC}(x_B) 
    & = \frac{1}{\rho_B}
    \int_0^{\frac{1-x_B}{x_B}Q^2}\hspace*{-.2cm} dm_j^2 J_2(m_j^2) \\
    & \times F_{3,5,6}^{(0)}\big(\xi(1+m_j^2/Q^2)\big) \\
  F_4^\text{JMC}(x_B)
    & = \int_0^{\frac{1-x_B}{x_B}Q^2}\hspace*{-.2cm} dm_j^2 J_2(m_j^2) 
    \Big\{ F_4^{(0)}\big(\xi(1+m_j^2/Q^2)\big) \\
    & + \frac{1-\rho_B}{\rho_B} F_5^{(0)}\big(\xi(1+m_j^2/Q^2)\big)
    \Big\} \ .
\end{split}
\end{align}

\section{Target mass corrections in the OPE formalism}
\label{app:GP}

We collect here for completeness the target mass corrections to
the electromagnetic structure functions obtained in the operator
product expansion formalism of De Rujula, Georgi and Politzer
\cite{Georgi:1976ve,DeRujula:1976tz}, see also
\cite{Schienbein:2007gr} for a thorough review and discussion: 
\begin{align}
  F_1^{GP}(x_B,Q^2) & 
    = \frac{x_B}{\rho_B} \Big[
      \frac{F_1^{(0)}(\xi,Q^2)}{\xi} + \frac{m_N^2 x_B }{Q^2 \rho_B}
      \Delta_2(x_B,Q^2) \Big]
    \nonumber \\
  F_2^{GP}(x_B,Q^2) & 
    = \frac{x_B^2}{\rho_B^3} \Big[
      \frac{F_2^{(0)}(\xi,Q^2)}{\xi^2} + 6 \frac{m_N^2 x_B }{Q^2 \rho_B}
      \Delta_2(x_B,Q^2) \Big]
    \nonumber \\
  F_L^{GP}(x_B,Q^2) & 
    = \frac{x_B}{\rho_B} \Big[
      \frac{F_L^{(0)}(\xi,Q^2)}{\xi} + 2 \frac{m_N^2 x_B }{Q^2 \rho_B}
      \Delta_2(x_B,Q^2) \Big]
    \label{eq:GP} 
\end{align}
where 
\begin{align}
  \Delta_2(x_B,Q^2) & = \int_\xi^1 dv 
    \Big[ 1 + 2\frac{m_N^2 x_B}{Q^2 \rho_B} (v-\xi)
    \Big] \frac{F_2^{(0)}(v,Q^2)}{v^2} \ ,
\end{align}
and $F_i^{(0)}$ are the perturbative structure functions computed in
the massless target approximation $m_N^2/Q^2 \ra 0$. Formulae for the
$F_{3-6}^{GP}$ structure functions can be found in
Ref.~\cite{Blumlein:1998nv}. 

Note that in the notation of Appendix~\ref{app:strfns}, differently
from Ref.~\cite{Schienbein:2007gr}, the longitudinal structure
function is defined such that 
\begin{align}
  F_L(x_B) = \frac{\rho^2}{2x_B} F_2(x_B) -  F_1(x_B) \ ,
  \label{eq:FLdef}
\end{align}
and
\begin{align}
  R(x_B) \equiv \frac{\sigma_L(x_B)}{\sigma_T(x_B)} 
    = \frac{F_L(x_B)}{F_1(x_B)}.
  \label{eq:Rdef}
\end{align}
This notation is explained in detail in the Appendices of
Ref.~\cite{Aivazis:1993kh}. Combining Eqs.~\eqref{eq:FLdef} and
\eqref{eq:Rdef} we obtain
\begin{align}
  F_1(x_B) = \frac{\rho_B^2}{2x_B} \frac{F_2(x_B)}{1+R(x_B)} 
\end{align}
in agreement with Ref.~\cite{Schienbein:2007gr}.

Equations~\eqref{eq:GP}
have been used to compute the OPE target mass corrections in
Figs.~\ref{fig:F2_TMC} and \ref{fig:R_TMC}.
Note that both $F_L^{GP}$ and $F_1^{GP}$ receive a correction
from an integral of $F_2^{(0)} \gg F_{1,L}^{(0)}$.
This explains the large size of the
target mass corrections for the OPE curves of Fig.~\ref{fig:R_TMC}.

\end{appendix}


\end{document}